\documentclass{dsfe}

\usepackage{txfonts}
\def\typeofarticle{Research article}
\def\currentvolume{4}
\def\currentissue{3}
\def\currentyear{2024}

\def\ppages{xxx--xxx.}
\def\DOI{10.3934/DSFE.2024016}
\def\Received{23 February 2024}
\def\Revised{25 June 2024}
\def\Accepted{09 July 2024}
\def\Published{ July 2024}
\usepackage{lineno}

\usepackage{graphicx}
\usepackage[skip=0.333\baselineskip]{caption}
\usepackage{subcaption}
\usepackage{microtype}
\usepackage[font=small,skip=-15pt]{caption}
\usepackage{amsmath}
\usepackage{amssymb}
\usepackage{amsfonts,mathtools}
\usepackage{amsthm}
\usepackage{float}
\usepackage[ruled]{algorithm2e}

\begin{document}
%\linenumbers

\title{\textls[-5]{A causal interactions indicator between two time series using extreme} variations in the first eigenvalue of lagged correlation matrices}

\author{%
  Alejandro Rodriguez Dominguez\affil{1,2} and Om Hari Yadav\affil{3}
}

% \shortauthors is used in copyright information in the end of the paper
\shortauthors{the Author(s)}

\address{%
  \addr{\affilnum{1}}{Department of Quantitative Research, Miralta Finance Bank S.A., Madrid, 28043, Spain}
  \addr{\affilnum{2}}{Department of Computer Science, University of Reading, Reading, United Kingdom}
  \addr{\affilnum{3}}{Department of Data Science, Indian Institute of Science Education and Research (IISER), Thiruvananthapuram, Kerala, India}}

% corresponding author
\corraddr{Email: arodriguez@miraltabank.com; Tel: +34620090488.
}

%\begin{abstract}
%A method to identify causal interactions between a pair of cause-and-effect variables using the variability of the explanatory power from the first eigenvalue is presented. The behavior of the largest eigenvalue for lagged correlation matrices follows a Tracy-Widom distribution, which can be derived from the homology of a Coulomb gas being compressed by two walls. Using this connection, causal interactions are defined as continuous pushing and pulling of the gas by the walls, which can be measured by the variability of the largest eigenvalue's explanatory power. Experiments validate the causal identification, both with controlled randomized trials and historical data. Financial market data is used to show the prediction capabilities of the indicator with respect to average stock return and realized volatility, making it a powerful and simple indicator for risk management. To the authors' knowledge it is the first time this indicator has been applied for this purpose. A liquidity risk management framework is presented with data from a brokerage business with the indicator being able to infer causal interactions and relationships in the liquidity business flow. The method can detect structural changes and breaks in time series, and causal interactions between a pair of cause-and-effect candidates variables, which can be monitored dynamically and not based solely on predictability, being more accurate than other existing methods well-established in the literature.

%\textbf{(200 to 300 words)}
%\end{abstract}

\begin{abstract}
This paper presents a method to identify causal interactions between two time series. The largest eigenvalue follows a Tracy-Widom distribution, derived from a Coulomb gas model. This defines causal interactions as the pushing and pulling of the gas, measurable by the variability of the largest eigenvalue’s explanatory power. The hypothesis that this setup applies to time series interactions was validated, with causality inferred from time lags. The standard deviation of the largest eigenvalue’s
explanatory power in lagged correlation matrices indicated the probability of causal interaction between time series. Contrasting with traditional methods that rely on forecasting or window-based parametric controls, this approach offers a novel definition of causality based on dynamic monitoring of tail events. Experimental validation with controlled trials and historical data shows that this method outperforms Granger’s causality test in detecting structural changes in time series. Applications to stock returns and financial market data show the indicator's predictive capabilities regarding average stock return and realized volatility. Further validation with brokerage data confirms its effectiveness in inferring causal relationships in liquidity flows, highlighting its potential for market and liquidity risk management.
\end{abstract}


\maketitle

\section{Introduction}
Time series correlations are widely used dependence metrics in finance and economics. The problem is that correlations are usually spurious, do not represent dependence for non-normal distributions, and heavily depend on the scale and frequency of the time series. Correlations are statistical measures that depend on the chosen time-window as point estimates, making them useless for dynamic dependencies and change point detection \citep{10.1093/ckj/sfab085, 10098076}. To avoid these pitfalls, statistical techniques such as copulas, entropy-based measures, network graphs, and parametric distance metrics have been applied to model dependence \citep{ inbookMarti}. These statistical methods, along with their flaws, are involved in finance and insurance for risk management, portfolio optimization, alpha generation, factor investing, credit lending, and more, with trillions of dollars depending on them every year. Causality could be one solution for measuring true dependencies, but at the cost of being difficult to measure and identify. Granger's economics Nobel Prize-winning causality test \citep{702ab909-8cb1-3c30-a5f1-ab4517d6cf1c} detects probabilistic causal relationships between time series using control variables. However, although it provides information about forecasting capabilities, it provides nothing about true causal interactions. The fact that it still remains one of the benchmarks for time series causal identification evidences the need for a statistical method using time series data that could detect causal interactions. \par

Causal relationships are based on interactions, which if strong enough, can modify the structure of the time series model. This work focuses on physical experiments with a clear definition for causal interactions and mapping statistical properties to time series causal identification. The Coulomb-Dyson gas of charges is a log-gas (or a two-dimensional Coulomb gas) consisting of repelling particles on the line in a global confining potential. An interaction can be defined as compressing the whole gas, which is equivalent to forcing the largest eigenvalue of a random matrix representing the gas inside the allowed band given by the Tracy-Widom distribution \citep{articleTracy}. This is the distribution for the large deviations of the first eigenvalue in a random matrix. The large deviation function is the excess free energy of the gas of eigenvalues forced to stay between two hard walls \citep{Cunden2018ThirdOrderPT}. Greater values in the first eigenvalue in a variogram can explain a third-phase gas transition pulled-to-pushed-type interaction. The first premise of this work is that a greater number of interactions implies greater probability of a true interaction. Finally, this system of a Coulomb-Dyson gas being pulled and compressed by two walls can be equivalent to a system of two pairs of cause-and-effect time series candidates. In this case, the large deviations of the first eigenvalue are analyzed for different lagged $2\times 2$ correlation matrices, with the lag applied to the causal candidate time series. The second premise of this work is that the interaction, detected in nature and by homology to the Coulomb-Dyson gas compression, must be causal due to the direction of time imposed by the lagged correlation matrices. Therefore, the interaction that is true with highest probability and in the direction of time must be causal. The standard deviation of the explanatory power of the first eigenvalue for all lags is computed as a causal interactions indicator. Large values of this indicator in a particular point in time means a high probability of a causal interaction, the interaction being measured by the extreme deviations of the first eigenvalue and the causality enforced by the direction of time in the lagged correlation matrices.\par

Based on these previous aspects, the main contributions proposed in
this paper follow:

\begin{itemize}
    \item A new method to detect causal interactions between two time series using random matrix theory (RMT) and eigenvalues' distributions being defined from physical experiments is presented, in the form of an indicator.
    
    \item To the best of the authors' knowledge, there is no causal identification method that focuses on random matrix theory (RMT) and eigenvalues' distributions. Some work uses RMT to search for structural changes in financial markets \citep{potters2005financial,articleJPB}. It differs from this paper in that they do not mention the variability of the variogram of the largest eigenvalue, but focus on its tendency and with no link to causal interactions. Whereas this work uses a different indicator consisting of the standard deviation of the variogram measured, in this case, by the explanatory power of the largest eigenvalue and not the eigenvalue itself to capture extreme variations, which maximize the probability of having a causal interaction.
    
    \item The method is able to capture structural changes in time series more dynamically than traditional econometric methods. It is also able to detect causal interactions in a pair of time series. In the experiments, the method is able to identify the causal variable from a pool of more than one thousand time series candidates. The same output is found with synthetic generated data.

    \item The indicator is relatively simple, easy for model validation of financial institutions and regulators, and can be used, as shown in the experiments, for risk management in flow business, including but not limited to liquidity risk in banks and brokerage units, foreign exchange flows in the global economy \citep{Mensi2021,Arkol2024}, order-book management \citep{Brown-1997} and high frequency trading \citep{Samsul-2019}. It is also useful for market risk analysis, regime-switch detection \citep{Balcilar2015}, structural breaks in the market \citep{Fenghua2017} and forecasting \citep{JingLi2021}.

    \item Finally, to the best of the authors' knowledge, there is no method that connects causality from an experimental physics point of view with causal interactions between time series \citep{Edinburgh.77.041108}.

\end{itemize}

The paper is organized as follows: in Section 2, a revision of the state-of-the-art method is presented; in Section 3, the Coulomb-Gas experimental setup analogy for causal interactions is described; in Section 4, the methodology for the indicator is shown; in Section 5, experiments are presented; and in Section 6, some conclusions are made and future work suggested.

% The Introduction section should provide a brief statement of the research background and whether the aim of the article was achieved. Please use the AIMS template to prepare your manuscript, before you submit to our journal. Please read carefully the instructions for authors at \url{http://www.aimspress.com} (Benoist (Year)). These are important instructions and explanations. References be cited in text in this way: By author name and year of publication in parentheses inside the punctuation (e.g., Finnegan et al., 2004; or Finnegan, 2004; or Finnegan and Fraser, 2004). 

\section{Literature review}
The main issue with applying causality to finance and economics is that it has many varied interpretations by authors in different fields \citep{Pearl09a,karimi2010generation,YARLAGADDA2003761, 10.1214/10-STS321}. A Nobel Prize-winning causal time series identification approach is the Granger causality test \citep{702ab909-8cb1-3c30-a5f1-ab4517d6cf1c}, which is not focused on empirical causality, and just considers linear prediction and time-lagged dependence between two time series. Despite being one of the most applied methods in history, it cannot identify true causal relationships as it does not take unknown confounders into account \citep{JohnCochrane1997}. The Granger test has been designed for the time and frequency domains, but the same issues persist \citep{Breitung2006-fe}. Causality has been studied with different frameworks, which are sometimes difficult to connect in quantum physics, statistical physics, and relativity theory \citep{Brukner2014-wy, Sklar2010-ny, Havas1969-gv}. This work focuses on the concept of causality from statistical physics.\par

\begin{figure}[h!]
\setlength\abovecaptionskip{0\baselineskip}
	\centering
	\includegraphics[width=100mm]{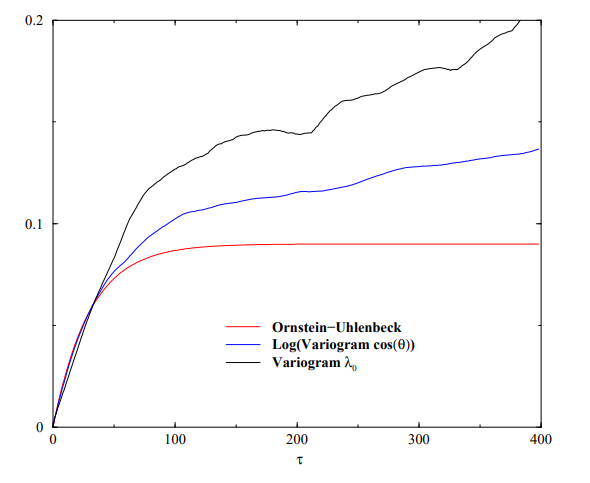}
	\caption{ Variogram of the top eigenvector from \citep{potters2005financial}.}
	\label{fig_descarga_4}
\end{figure}

Causality is defined as the requirement that two points in space-time cannot communicate with each other if they are separated by a spacelike distance \citep{Rosenfelder1989}. Therefore causality implies interactions. Random matrix theory has been applied for detecting structural changes in financial market dynamics, in particular the variogram of the largest eigenvalue \citep{potters2005financial, articleJPB}. An expression for the variogram of the top eigenvalue is obtained in \citep{potters2005financial}:
\begin{equation}
   \left|\mathrm{\lambda}_{\mathrm{t+\tau}}\mathrm{-} \mathrm{\lambda}_\mathrm{t}\right|^\mathrm{2}=\mathrm{2\epsilon}\left(\mathrm{1-exp}\left(\mathrm{-\epsilon\ \tau}\right)\right)
   \label{eqvario}
\end{equation}
This result is expected when a "true" dynamical evolution of the structure of the matrix is absent. These results assume that the correlation matrix is time independent. Any significant deviation from (\ref{eqvario}) would indicate a genuine evolution of the market structure. Figure \ref{fig_descarga_4} shows a real evolution of the strength of the average correlation with time. Finally, the non-stationary effects in financial markets are investigated by studying the dynamics of the top eigenvalue and eigenvector. If the true top eigenvector (or eigenvalue) did not evolve in time, the variograms in (\ref{eqvario}) should converge to their asymptotic limits. If the correlations' structure changes over extended periods of time, which would lead to an increased asymptotic value for the variograms including a long time tail. The asymptotic value of these variograms is furthermore much larger than the theoretical one. This non-stationarity of financial markets is expected and dangerous for out-of-sample risk management \citep{potters2005financial}. In Figure \ref{fig_descarga_4}, the variograms display the existence of genuine fluctuations of the market mode on a time scale of 100 days, superimposed to the initial noise-dominated regime. These fluctuations, when becoming extreme, can be captured by the indicator presented in this paper and related to causal interactions. \par

The Tracy-Widom distribution of the largest eigenvalue for extreme values has been studied for random dynamical systems \citep{Majumdar2013}. In particular, the Coulomb gas analogy, which is used in this work for causal interactions in pairs of time series, has been applied before to derive the deviation functions describing the tails of the distribution of the largest eigenvalue \citep{Hasting1980, Nadal2011}.

\section{Coulomb gas analogy for detecting causal interactions}
Random matrix theory analyzes the eigenvalue spectra of random matrices, and in particular the Tracy-Widom region is focused on the edges. The distribution of eigenvalues of a $N\times N$ Hermitian matrix, with independent and identically distributed (i.i.d.) upper triangular entries, converges in probability as $N\rightarrow\infty$ to the Wigner semicircle distribution. The probability distribution of the largest eigenvalue, $\lambda_{max}$, is more difficult to estimate. It has been shown that the standard deviation of this distribution scales to $N^{-1/6}$. Tracy-Widom revealed that for large $N$, the distribution of a random variable $\chi$ is independent of $N$: $P(\chi \leq x)=F_{\beta}(x)$, which is called the Tracy-Widom distribution and depends on $\beta$. For $\beta=2$:
\begin{equation}
 F_2\left(x\right)=exp\left(-\int_{x}^{\infty}{\left(z-x\right)q^2\left(z\right)}dz\right)
\end{equation}
Here, $q(x)$ satisfies the special case of $\alpha = 0$ of the Painleve II equation:
\begin{equation}
q^{\prime\prime}(z)=zq(z)+2q^3(z)+\alpha
\end{equation}
For $\alpha = 0$, the solution only requires the right-tail boundary condition for its unique specification: $q(z)\sim Ai(x)$ as $z\rightarrow\infty$. $Ai$ is the Airy function
\begin{equation}
Ai(x)=\frac{1}{\pi}\int_{0}^{\infty}\cos{\left(\frac{t^3}{3}+xt\right)dt}
\end{equation}
which satisfies the differential equation $Ai^{\prime\prime}(z)-z Ai(z)=0$ and vanishes like the following expression: $Ai(z) \sim (1/2\sqrt{\pi}z^{1/4})\exp^{-(2/3)z^{3/2}}$ as $z\rightarrow\infty$ \citep{Nadal2011,Hasting1980}. 
For large deviations of the largest eigenvalue, the probability of $\lambda_{max}$ for large $N$ is \citep{Nadal2011}:
\begin{equation}
    \mathcal{P}(\lambda_{max}=t)\approx\left\{\begin{matrix}exp\left\{-\beta N^2\psi_{-}\left(\frac{t}{\sqrt N}\right)+\dots\right\}&for\ t<\sqrt{2N\ }\ and\ \left|t-\sqrt{2N\ }\right|\approx O(\sqrt N)\\\frac{1}{a_{\beta}N^{-1/6}}F_\beta^{\prime}\left(\frac{t-\sqrt{2N\ }}{a_{\beta}N^{-1/6}}\right)&for\ \left|t-\sqrt{2N\ }\right|\approx O(N^{-1/6})\\exp\left\{-\beta N\psi_{+}\left(\frac{t}{\sqrt N}\right)+\dots\right\}&for\ t>\sqrt{2N\ }\ and\ \left|t-\sqrt{2N\ }\right|\approx O(\sqrt N)\\\end{matrix}\right.
\end{equation}
$F_\beta^{\prime}$ is the Tracy–Widom distribution, $\psi_{-}$ and $\psi_{+}$ are the left and right large deviation functions describing the tails of the distribution of $\lambda_{max}$, and both were computed by simple physical methods exploiting the Coulomb gas analogy \citep{Hasting1980}.\par
\begin{figure}[h!]
\setlength\abovecaptionskip{0\baselineskip}
	\centering
	\includegraphics[width=100mm]{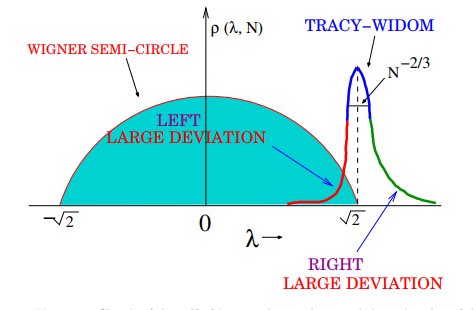}
	\vspace{0.4cm}
	\caption{The Tracy-Widom distribution from \citep{Satya2020}.}
	\label{figure2}
\end{figure}
\begin{figure}[h!] \setlength\abovecaptionskip{0\baselineskip}
	\centering
	\includegraphics[width=100mm]{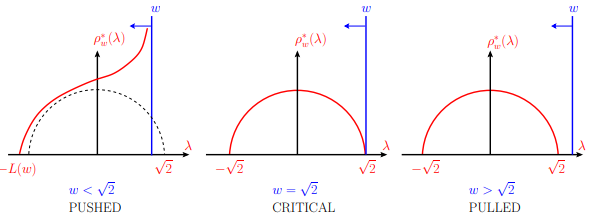}
	\vspace{0.4cm}
	\caption{The effect of the presence of a wall on the Wigner semi-circle.}
	\label{figure3}
\end{figure}

The physics of the left and right tail are very different. In the case of the left tail, the semi-circular charge density of the Coulomb gas is pushed by the hard wall, resulting in a reorganisation of all of the $N$ charges, which creates an energy difference. In contrast, for the right tail, the dominant fluctuations are caused by pulling a single charge away (to the right) from the Wigner sea (the blue area from Figure \ref{figure2}), leading to an energy difference of $O(N)$. The different behavior of the probability distribution leads to a phase transition. In Figure \ref{figure3}, the effect of the presence of a wall on the Wigner semi-circle is shown. If $N$ is large but finite, the Tracy-Widom distribution describes the crossover function between stable and unstable regimes of a random dynamical system \citep{Majumdar2013}, and moreover, there is a third-order phase transition between these regimes, which also takes place in the Coulomb gas, two-dimensional Quickest Change Detection, and several other systems of interest. \par

\section{Causal interactions time series indicator}
Given $\boldsymbol{X}, \boldsymbol{Y}\in\mathbb{R}^{L\times 1}$, a pair of cause-and-effect variables in the form of time series of length $L$. The lagged correlation matrix is computed with the returns of $\boldsymbol{X}$ and $\boldsymbol{Y}$, $\boldsymbol{r_X}, \boldsymbol{r_Y}\in\mathbb{R}^{L\times 1}$, and lagging one of them by $\tau$ time-steps of time $t$. For the relationship to be causal, the direction of time is such that the lagged variable must be the causal candidate time series $\boldsymbol{X}$. The lagged correlation coefficient is:
\begin{equation}
    \mathcal{C}_{XY}(\tau)=\langle \boldsymbol{r_X}^t,\boldsymbol{r_Y}^{t+\tau}\rangle
    \label{corrlag}
\end{equation}
such that $\mathcal{C}_{XY}(\tau=0)=\mathcal{C}_{XY}$ is the standard correlation coefficient. Whereas the correlation matrix of the standard case is symmetric, the lagged case $\boldsymbol{R_{XY}}=\left[\begin{matrix}1&C_{YX}(\tau)\\C_{XY}(\tau)&1\\\end{matrix}\right]$ is non-symmetric.

The indicator for causal interactions, $\sigma_{\Gamma}$, is defined as the standard deviation of the explanatory power for the first eigenvalue, $\Gamma=\frac{\lambda_1}{\lambda_1+\lambda_2}$,  in $2\times 2$ lagged correlation matrices for different values of lag $\tau=1,\dots \max\tau$. The indicator at time $t$ is computed as follows:
\begin{itemize}
    \item For each lag $\tau$, the lagged correlation matrix is computed as $\mathcal{C}_{XY}(\tau) = \langle \boldsymbol{r_X}^{t-\tau}, \boldsymbol{r_Y}^{t} \rangle$, with the lag applied to the candidate causal time series returns $\boldsymbol{r_X}$, while the effect $\boldsymbol{r_Y}$ remains fixed for all lags.
    \item A principal component analysis is performed on each lagged $\mathcal{C}_{XY}(\tau)$, and the explanatory power of the first eigenvalue is recorded as $\Gamma(\tau) = \frac{\lambda_1}{\lambda_1 + \lambda_2}$.
    \item The same process is repeated for all lags, $\tau=1,\dots \max\tau$, with $\max\tau$ being a model hyperparameter. Finally, the standard deviation of all $\Gamma(\tau)$ values, for $\tau=1,\dots \max\tau$ gives the indicator at time $t$,  $\sigma_{\Gamma}(t)$:
    \begin{equation}
    \sigma_{\Gamma}(t)=Std\left(\Gamma(1),\dots,\Gamma(\max\tau)\right)
    \end{equation}
\end{itemize}

Higher values of $\sigma_{\Gamma}(t)$, $\forall t$, means a higher probability of having a causal interaction from $\boldsymbol{X}$ to $\boldsymbol{Y}$. In Algorithm \ref{alg:eigen}, the method is presented.
%, , instead of $\lambda_1-\lambda_2$ which is used for the standard deviation of the Tracy-widom distribution $\sigma_{n\beta}$. 
%This allows us to deal with a measure with empirical sense for the method. With extreme values of $\sigma_{\Gamma}$ indicating deviations from the Tracy-Widom distribution $F_\beta$ implying asymptotically independence versus structural time (causal) relationships. 

\begin{algorithm}[h!]
\caption{Causal interactions time series indicator for 2-variables set ($\boldsymbol{\sigma_{\Gamma}}$)}
\label{alg:eigen}
\KwData{Causal time series candidate $\boldsymbol{X}\in\mathbb{R}^{L\times 1}$ and effect candidate time series $\boldsymbol{Y}\in\mathbb{R}^{L\times 1}$, both of length $L$; lag values, $\tau=1,\dots \max\tau$, for a given $\max\tau$; $k=t,\dots,T$ are the timestamps for the time series of the indicator.}
\KwResult{The time series $\boldsymbol{\sigma_{\Gamma}(t\rightarrow T)}=\left\{\sigma_{\Gamma}(t),\dots,\sigma_{\Gamma}(T)\right\}$, with $k=t,\dots,T$.}
\For{$t \le k \leq T$}{    
    $\boldsymbol{X}(k) = \boldsymbol{X}[k-L\colon k]$;\  \ $\boldsymbol{X}(k) = (\boldsymbol{X}(k) -mean(\boldsymbol{X}(k)))/std(\boldsymbol{X}(k)) $ \\
    $\boldsymbol{Y}(k) = \boldsymbol{Y}[k-L\colon k]$;\  \ $\boldsymbol{Y}(k) = (\boldsymbol{Y}(k) -mean(\boldsymbol{Y}(k)))/std(\boldsymbol{Y}(k)) $ \\
    \For{$1 \le \tau \leq \max\tau$}{
        $\boldsymbol{X}(k,\tau)=\boldsymbol{X}(k).shift(\tau)$\\
        $PCA(k,\tau) = PCA([\boldsymbol{Y}(k),\boldsymbol{X}(k,\tau)]^T)$\\
        $V(k,\tau)=\frac{\lambda_1(k,\tau)}{\lambda_1(k,\tau)+\lambda_2(k,\tau)}$\\
    }
$\sigma_{\Gamma}(k)= Std(V(k,1),\dots,V(k,\max\tau))$\\
$\boldsymbol{\sigma_{\Gamma}(t\rightarrow T)}.append(\sigma_{\Gamma}(k))$

}
\end{algorithm}

\section{Experiments and discussion}
\subsection{Datasets description}
\label{secdesc}
Two sets of experiments were performed. The first one uses financial market daily time series data sourced from Bloomberg, spanning the years 2013 to 2023. The second set of experiments uses daily time series data from end-of-day files of a brokerage business containing all clients' cash and trading activities, flows, and positions. The anonymous dataset was provided by Miralta Finance Bank S.A. A description of the datasets is given next:
\begin{itemize}
    \item For the first set of experiments in Section 5.3, a financial market dataset consisting of 110 daily time series of stocks' causal drivers candidates was used. These are enumerated in Table \ref{Driverslist}, and consist of daily time series of pricing, macroeconomic, products and market indicators data from the following asset classes: credit, rates, commodities and equities indexes, FX spot rates (X-rate and spot is the same), commodities, equities and government bonds futures' indexes, credit default swap (CDS) indexes and implied volatility data. A second dataset consists of effect time series candidates, daily prices for twelve SP500 stocks from different sectors. Notable examples include technology sector stocks such as Apple, consumer discretionary sector stocks like McDonald's Corporation, energy sector stocks such as Schlumberger Limited, and financial sector stocks like Goldman Sachs Group, Inc. 
    
    \item For the second set of experiments in Section 5.4, brokerage business variables are included, which comprise data for approximately 15,000 clients (as detailed in Table \ref{brokeragevariables}). This dataset provides information on various aspects such as trading volume data, cash-flow data, account statuses, and trade executions. This offers valuable insights into brokerage operations, client behavior and liquidity analysis for an institution like a bank, which must be managed optimally for regulatory purposes.
\end{itemize}

For the first dataset, the goal is to use the indicator, $\sigma_{\Gamma}$, to measure causal interactions and relationships between causal drivers' candidates and stocks from the SP500. The second dataset is focused on finding causal relationships via interactions between a brokerage business and its liquidity levels. But first, in the next section, a randomized experiment is performed.

\begin{figure}[t]
\setlength\abovecaptionskip{0\baselineskip}
	\centering
	\includegraphics[width=100mm]{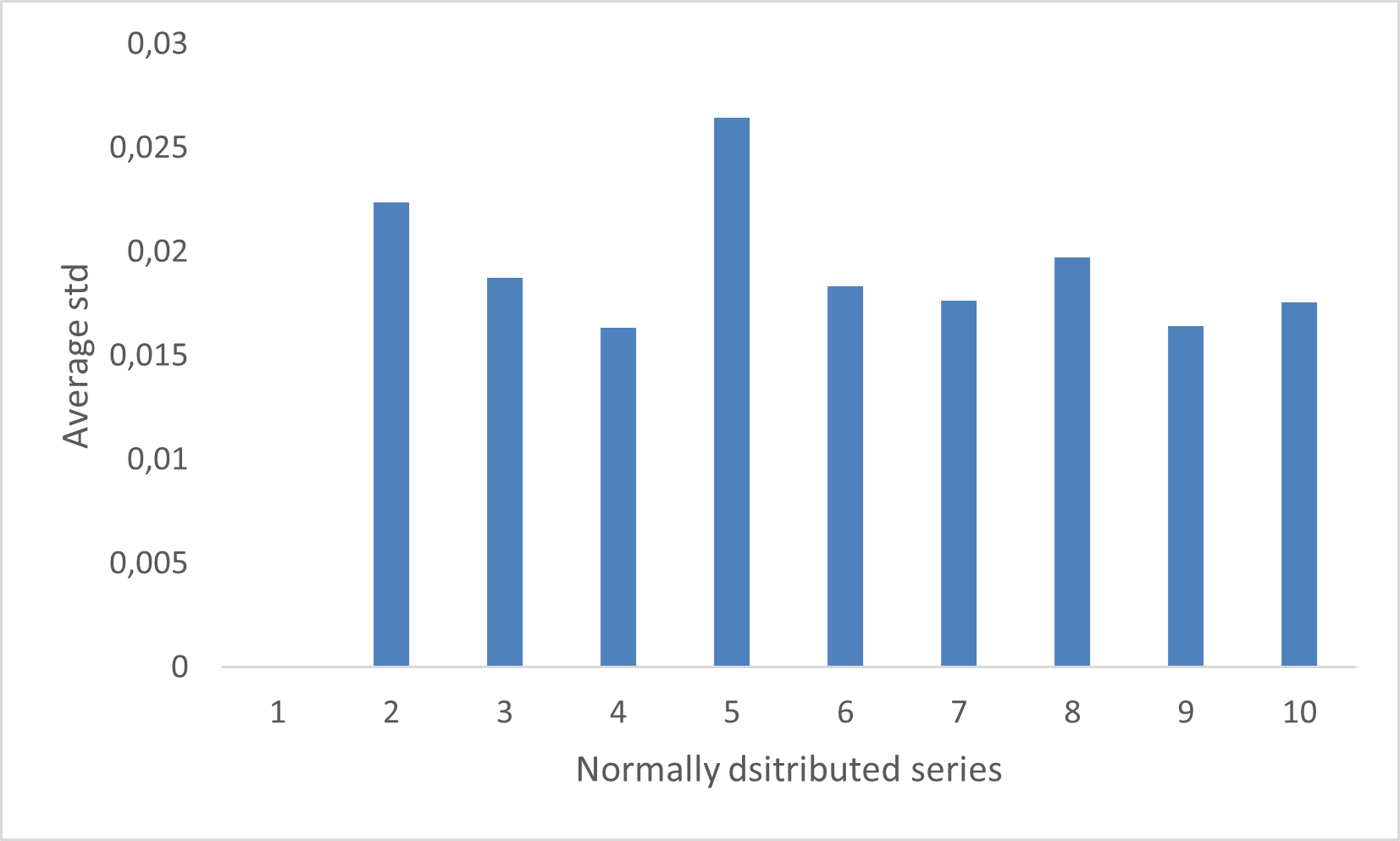}
	\vspace{0.4cm}
	\caption{Average of $\sigma_{\Gamma}$ for 10 i.i.d. time series generated from a normal distribution (400 time-stamps generated).}
	\label{figure_10Norm}
	
\vspace{0.4cm}
\setlength\abovecaptionskip{0\baselineskip}
	\centering
	\includegraphics[width=100mm]{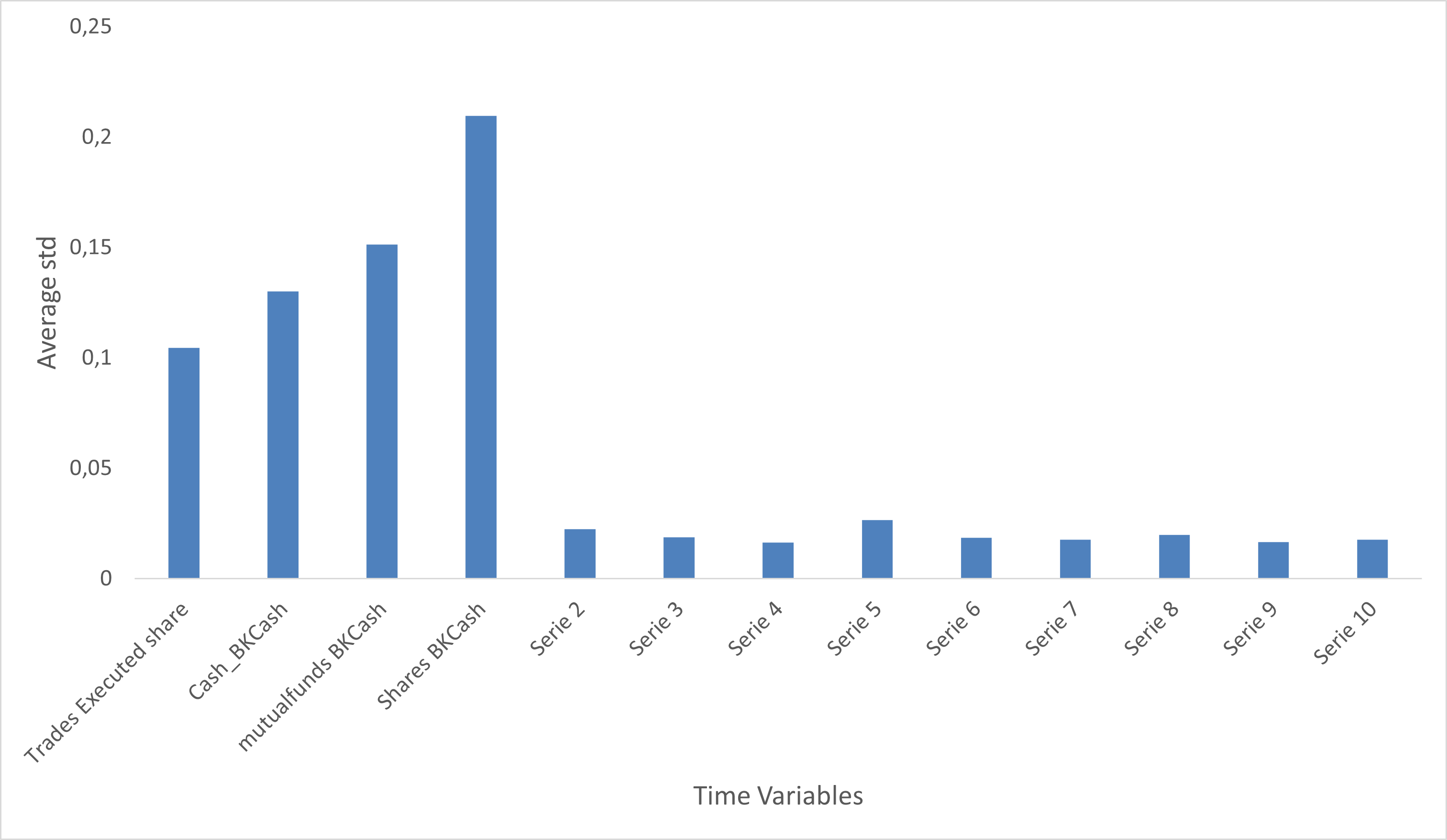}
	\vspace{0.4cm}
	\caption{Average of $\sigma_{\Gamma}$, the same case as in Figure \ref{figure_10Norm}. In this case, together with the average $\sigma_{\Gamma}$ for pairs of time series from the brokerage dataset of the experiments (with known causal interactions and structural changes). }
	\label{figure_10Norm_causes}
\end{figure}

\subsection{A Randomized experiment for structural change detection}
Figure \ref{figure_10Norm} shows the average of $\sigma_{\Gamma}$ for 10 i.i.d. time series generated from a normal distribution. For this, Algorithm \ref{alg:eigen} is applied to sets of 2 different time series. In Figure \ref{figure_10Norm_causes}, the average values of $\sigma_{\Gamma}$ from the i.i.d. series' previous case are compared with the average values of Algorithm \ref{alg:eigen} applied to time series variables from the brokerage dataset, which was described in the previous section, and includes true causal interactions and structural changes, as prior knowledge for the experiments. The hypothesis that the indicator captures structural changes can be simply verified from this experiment, as for i.i.d. generated time series in Figure \ref{figure_10Norm}, and the values of the cumulative sum of the indicator $\sigma_{\Gamma}$ are much smaller than the values for time series with structural relationships and changes in Figure \ref{figure_10Norm_causes}.

\subsection{Causal interactions market risk indicator}
The first set of experiments tests the ability of the indicator in detecting structural changes in the effect candidate variable. For that, the change is synthetically forced. For the different stocks, autoregressive moving average ARMA models are calibrated to historical time series returns data. At a specific and predefined point in time (03/10/2018), the ARMA parameters are forced to change in different ways, and the ARMA models with this new set of parameters are calibrated with the rest of the historical time series data, from that point onwards.
\begin{figure}[h!]
\begin{subfigure}{.475\linewidth}
  \includegraphics[width=\linewidth]{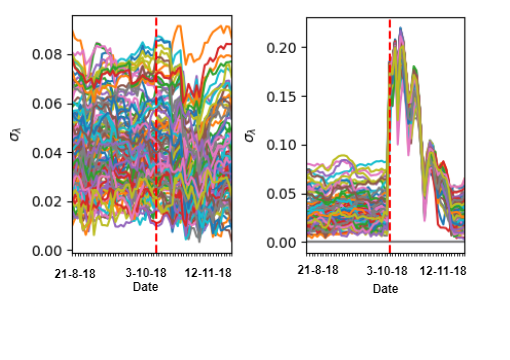}
  \caption{Goldman Sachs}
  \label{MLEDdet}
\end{subfigure}\hfill % <-- "\hfill"
\begin{subfigure}{.475\linewidth}
  \includegraphics[width=\linewidth]{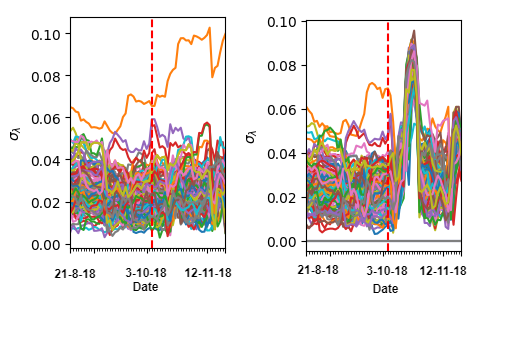}
  \caption{Apple}
  \label{energydetPSK}
\end{subfigure}

\medskip % create some *vertical* separation between the graphs
\begin{subfigure}{.475\linewidth}
  \includegraphics[width=\linewidth]{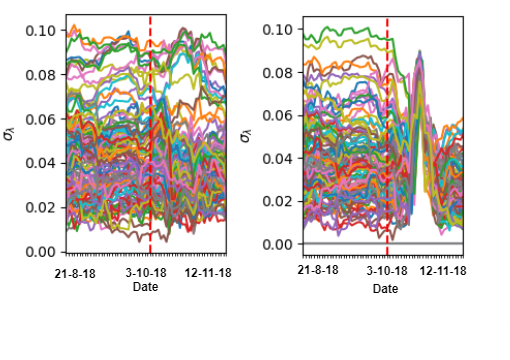}
  \caption{Schlumberger}
  \label{velcomp}
\end{subfigure}\hfill % <-- "\hfill"
\begin{subfigure}{.475\linewidth}
  \includegraphics[width=\linewidth]{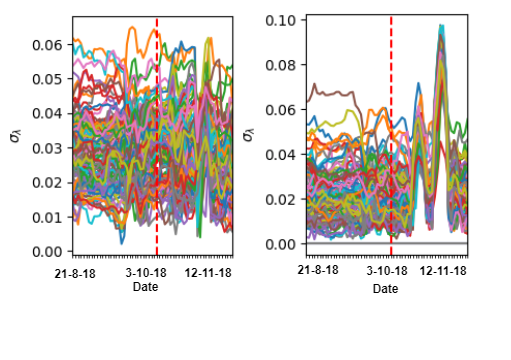}
  \caption{McDonald's}
  \label{estcomp}
\end{subfigure}
\medskip
\medskip
\vspace{0.4cm}
\caption{time series of the indicator $\sigma_{\Gamma}$ for all driver candidates and four SP500 stocks \textls[-20]{as effect variables. A structural change with calibrated ARMA models are forced for the effect variables at 03/10/2018 for all cases. Right sub-figures include the change, and left ones do not.}}
\label{figARMA}
\end{figure}

 In Figure \ref{figARMA}, a pair of figures are shown for four stocks: the figures on the left are the cases where the ARMA parameters do not change on that particular test day, and the right figures are the ones with changes in the ARMA parameters. In these figures, all of the cases using the four stocks as the effect variable and the 110 causal drivers' candidates from Table \ref{Driverslist} as causal variables are shown. It is clearly seen how the indicator $\sigma_{\Gamma}$, computed for all pairs of stock/drivers' candidates, dramatically jumps. This is what was expected for this indicator, the ability to accurately capture structural changes in time series. In the left figures in Figure \ref{figARMA}, indicators have higher values for the candidates with more chances to be related by causal interactions; and in the right figures, at the change point, this distance can have a reduced appearance. However, looking at the scale of the axis, it is clear that the gap between the indicators for different drivers is increased, and this is caused by the jump of all of the indicators which changes the scale. Even though all indicators jump at the change point, the causal drivers of the change have higher indicator values post-change. The structural changes have been identified, but there is still the need to test the indicator for identifying causal interactions with the extreme values of the indicator.

%\begin{table}[H]
%\begin{center}
%\caption{Caption of the table. Tables should be placed in the main text near to the first time they are cited.}
%\begin{tabular}{ccc} \hline
% & & \\\hline
% & & \\
% & & \\
% & & \\
% & & \\
% & & \\
% & & \\
% & & \\
% & & \\
% & & \\\hline
% & Note:(Table body should be created by MS word table function; three-line table is preferred.)
%\end{tabular}
%\end{center}
%\end{table}

\begin{table}[h!]
\begin{center}
\caption{List of financial markets' variables (cause-and-effect candidates) with time series from Bloomberg as described in Section \ref{secdesc}.}
\vspace{5mm}
\resizebox{\textwidth}{!}{%
\begin{tabular}{cccc}\hline
Ticker                  & Description                     & Ticker                     & Description                         \\
1-3 Year EU             & 1-3 Year EU Swap rate           & Generic 1st 'US' Future    & US 30-Year T-Bond Futures           \\
3-5 Years EU            & 3-5 Years EU Swap rate          & Generic 1st 'Z ' Future    & FTSE 100 Index Futures              \\
AUD-JPY X-RATE          & AUD-JPY X-RATE                  & Global Aggregate           & All World Credit Full Spectrum      \\
Argentine Peso Spot     & Argentine Peso Spot             & Global High Yield          & All world High Yield Credit         \\
Australian Dollar Spot  & Australian Dollar Spot          & HANG SENG INDEX            & HANG SENG INDEX                     \\
BRAZIL IBOVESPA INDEX   & BRAZIL IBOVESPA INDEX           & IBEX 35 INDEX              & IBEX 35 INDEX                       \\
Brazilian Real Spot     & Brazilian Real Spot             & Indian Rupee Spot          & Indian Rupee Spot                   \\
British Pound Spot      & British Pound Spot              & Japanese Yen Spot          & Japanese Yen Spot                   \\
CAC 40 INDEX            & CAC 40 INDEX                    & LME COPPER    3MO (\$)     & LME COPPER    3MO (\$)              \\
CAD-CHF X-RATE          & CAD-CHF X-RATE                  & LME COPPER   SPOT (\$)     & LME COPPER   SPOT (\$)              \\
CAD-JPY X-RATE          & CAD-JPY X-RATE                  & MSCI INDIA                 & India cross asset index             \\
CHF-JPY X-RATE          & CHF-JPY X-RATE                  & Mexican Peso Spot          & Mexican Peso Spot                   \\
Canadian Dollar Spot    & Canadian Dollar Spot            & NIKKEI 225                 & NIKKEI 225                          \\
Cboe Volatility Index   & Cboe Volatility Index           & New Zealand Dollar Spot    & New Zealand Dollar Spot             \\
Corporate               & All world Corporate Credit         & Norwegian Krone Spot       & Norwegian Krone Spot                \\
Credit                  & All world Credit full spectrum     & Pan-European High Yield    & Pan-European High Yield Credit      \\
DAX INDEX               & DAX INDEX                       & Polish Zloty Spot          & Polish Zloty Spot                   \\
DOLLAR INDEX SPOT       & DOLLAR INDEX SPOT               & RF Global Inflation-Linked & Fixed Income Inflation linked bonds \\
EM USD Aggregate        & Emerging Markets USD Aggregate  & RF Global Treasury         & Fixed Income Global Goverment bonds \\
EUR-AUD X-RATE          & EUR-AUD X-RATE                  & S\&P/BMV IPC               & S\&P/BMV IPC                        \\
EUR-CAD X-RATE          & EUR-CAD X-RATE                  & S. African Rand Spot       & S. African Rand Spot                \\
EUR-GBP X-RATE          & EUR-GBP X-RATE                  & Singapore Dollar Spot      & Singapore Dollar Spot               \\
EUR-JPY X-RATE          & EUR-JPY X-RATE                  & South Korean Won Spot      & South Korean Won Spot               \\
EUR-NOK X-RATE          & EUR-NOK X-RATE                  & Swedish Krona Spot         & Swedish Krona Spot                  \\
EUR-PLN X-RATE          & EUR-PLN X-RATE                  & Swiss Franc Spot           & Swiss Franc Spot                    \\
EUR-SEK X-RATE          & EUR-SEK X-RATE                  & TR/CC CRB ER Index         & Commodities Futures Index Price     \\
Euro Spot               & Euro Spot                       & Turkish Lira Spot          & Turkish Lira Spot                   \\
FTSE 100 INDEX          & FTSE 100 INDEX                  & U.S. Aggregate             & U.S. credit full spectrum           \\
FTSE MIB INDEX          & FTSE MIB INDEX                  & U.S. Corporate High Yield  & U.S. Corporate High Yield           \\
GBP-AUD X-RATE          & GBP-AUD X-RATE                  & U.S. Treasury              & US Treasury index                   \\
GBP-CAD X-RATE          & GBP-CAD X-RATE                  & UK Gilts 10 Yr             & UK Gilts 10 Yr                      \\
Generic 1st 'CF' Future & CAC 40 Futures                  & UK Gilts 30 Year           & UK Gilts 30 Year                    \\
Generic 1st 'CO' Future & Brent Crude Oil Futures ICE     & US Generic Govt 10 Yr      & 10 Yr Treasury                      \\
Generic 1st 'FV' Future & US 5-Year T-Note Futures        & US Generic Govt 2 Yr       & 2 Yr Treasury                       \\
Generic 1st 'GX' Future & DAX Index Futures               & US Generic Govt 5 Yr       & 5 Yr Treasury                       \\
Generic 1st 'HG' Future & Copper High Grade Futures COMEX & USD-CAD X-RATE             & USD-CAD X-RATE                      \\
Generic 1st 'HO' Future & Heating Oil \#2 Futures NYMEX   & USD-CHF X-RATE             & USD-CHF X-RATE                      \\
Generic 1st 'KC' Future & Coffee “C” Futures              & USD-GBP X-RATE             & USD-GBP X-RATE                      \\
Generic 1st 'QS' Future & Low Sulphur Gasoil Futures ICE  & USD-JPY X-RATE             & USD-JPY X-RATE                      \\ 
Generic 1st 'TY' Future & US 10-Year T-Note Futures \\\hline  
\label{Driverslist}
\end{tabular}
}
\end{center}
\end{table}

\begin{figure}[hbt!]
\centering
\begin{subfigure}{.23\linewidth}
  \includegraphics[width=\linewidth]{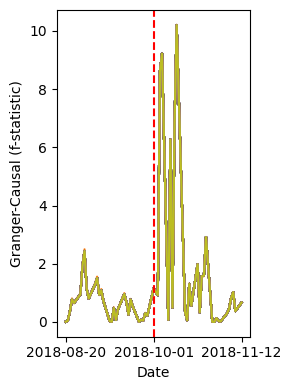}
  \caption{Goldman Sachs}
  \label{MLEDdet}
\end{subfigure}\hfill % <-- "\hfill"
\begin{subfigure}{.23\linewidth}
  \includegraphics[width=\linewidth]{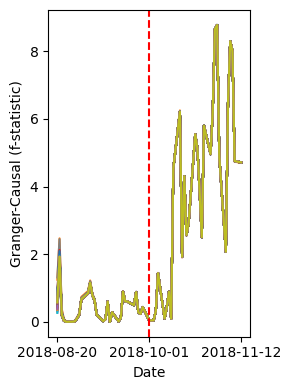}
  \caption{Apple}
  \label{energydetPSK}
\end{subfigure}\hfill % <-- "\hfill"
\begin{subfigure}{.23\linewidth}
  \includegraphics[width=\linewidth]{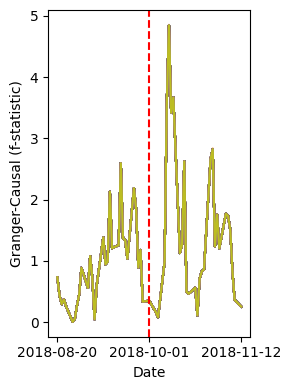}
  \caption{Schlumberger}
  \label{velcomp}
\end{subfigure}\hfill % <-- "\hfill"
\begin{subfigure}{.23\linewidth}
  \includegraphics[width=\linewidth]{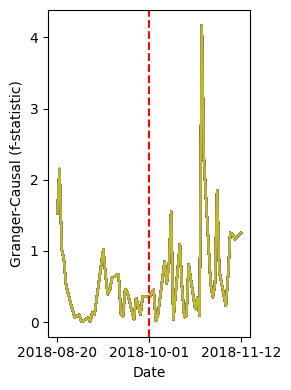}
  \caption{McDonald's}
  \label{estcomp}
\end{subfigure}\\
\captionsetup{skip=10pt} % Add space between subfigure captions and main caption
\caption{time series analysis presents the Granger causality test f-statistic for all driver candidates, indicating overlapping values across all causal candidates. The study encompasses four S\&P 500 stocks, scrutinizing them as effect variables. Additionally, a structural change is imposed on the effect-variables on January 10, 2018, employing calibrated ARMA models across all cases.
}
\label{figARMAGCT}
\end{figure}

Next, the same experiment shown in Figure \ref{figARMA} for the case of the causal indicator $\sigma_{\Gamma}$ is performed for the Granger causality test. In this case, a rolling test is computed for the same time period than in Figure \ref{figARMA}, for all causal drivers' candidates and the same four SP500 stocks. The forced change in ARMA structure is done at 01/10/2018, for all cases in these Granger causal test experiments. Results are shown in Figure \ref{figARMAGCT}, where the time series of the f-statistic for the Granger causality test (rolling-window-based) are shown for all cases. It can be seen that the Granger causality test captures the date of the structural changes, but with less precision than the proposed causal interactions' indicator $\sigma_{\Gamma}$. Also, the Granger test has an overlap for all the causal candidates' f-statistic time series values, in contrast to the $\sigma_{\Gamma}$ indicator, which can differentiate between ARMAs adjusted to different causal candidates' variables. This means that the $\sigma_{\Gamma}$ indicator has more precision detecting the date of the structural change for all cases than the Granger test. Also, the indicator $\sigma_{\Gamma}$ contains more information about the structural changes and the underlying time series processes, as it shows different values depending on the causal candidates' time series variables used to fit the ARMA models.

In Table \ref{statisticsstocks}, the KS and Anderson-Darling statistics, p-values, and the Wasserstein distance are shown for the four stocks presented in the experiments, in order to validate that all come from different distributions, enriching the generalization of the results. Due to space constraints, results are shown for these four stocks only, although original experiments were performed with twelve of them in all cases.

\begin{table}[h!]
\centering
\caption{Statistics for the four SP500 stocks shown in the experiments.}
\label{statisticsstocks}
\vspace{5mm}
\resizebox{\textwidth}{!}{%
\begin{tabular}{llrrrr}\hline
\textbf{List 1} & \textbf{List 2} & \multicolumn{1}{l}{\textbf{KS Statistic}} & \multicolumn{1}{l}{\textbf{KS P-Value}} & \multicolumn{1}{l}{\textbf{Anderson-Darling Statistic}} & \multicolumn{1}{l}{\textbf{Wasserstein Distance}} \\ \hline
SCHLUMBERGER LTD & APPLE INC & 0.0872 & 0.0000 & 217.4060 & 0.0042 \\
SCHLUMBERGER LTD & GOLDMAN SACHS GROUP INC & 0.0658 & 0.0000 & 196.3330 & 0.0043 \\
SCHLUMBERGER LTD & MCDONALDS CORP & 0.1523 & 0.0000 & 1,163.2190 & 0.0084 \\
APPLE INC & GOLDMAN SACHS GROUP INC & 0.0372 & 0.0655 & 12.2670 & 0.0010 \\
APPLE INC & MCDONALDS CORP & 0.1212 & 0.0000 & 519.5290 & 0.0044 \\
MCDONALDS CORP & GOLDMAN SACHS GROUP INC & 0.1082 & 0.0000 & 524.3530 & 0.0042 \\ \hline
\end{tabular}%
}
\begin{flushleft}
Note: This table displays the statistical results of Kolmogorov-Smirnov (KS) tests for data similarity, Anderson-Darling tests for non-normal data distribution, and Wasserstein distances for measuring the distance between probability distributions. It is computed between pairs of stock return distributions for selected companies from different sectors including significance level. Significant p-values ($\leq$ 0.05) indicate notable differences in distributions, while larger test statistics signify greater dissimilarity.
\end{flushleft}
\end{table}

\begin{figure}[h!]% [H] is so declass\'e!

\begin{subfigure}{0.45\textwidth}
\includegraphics[width=\textwidth]{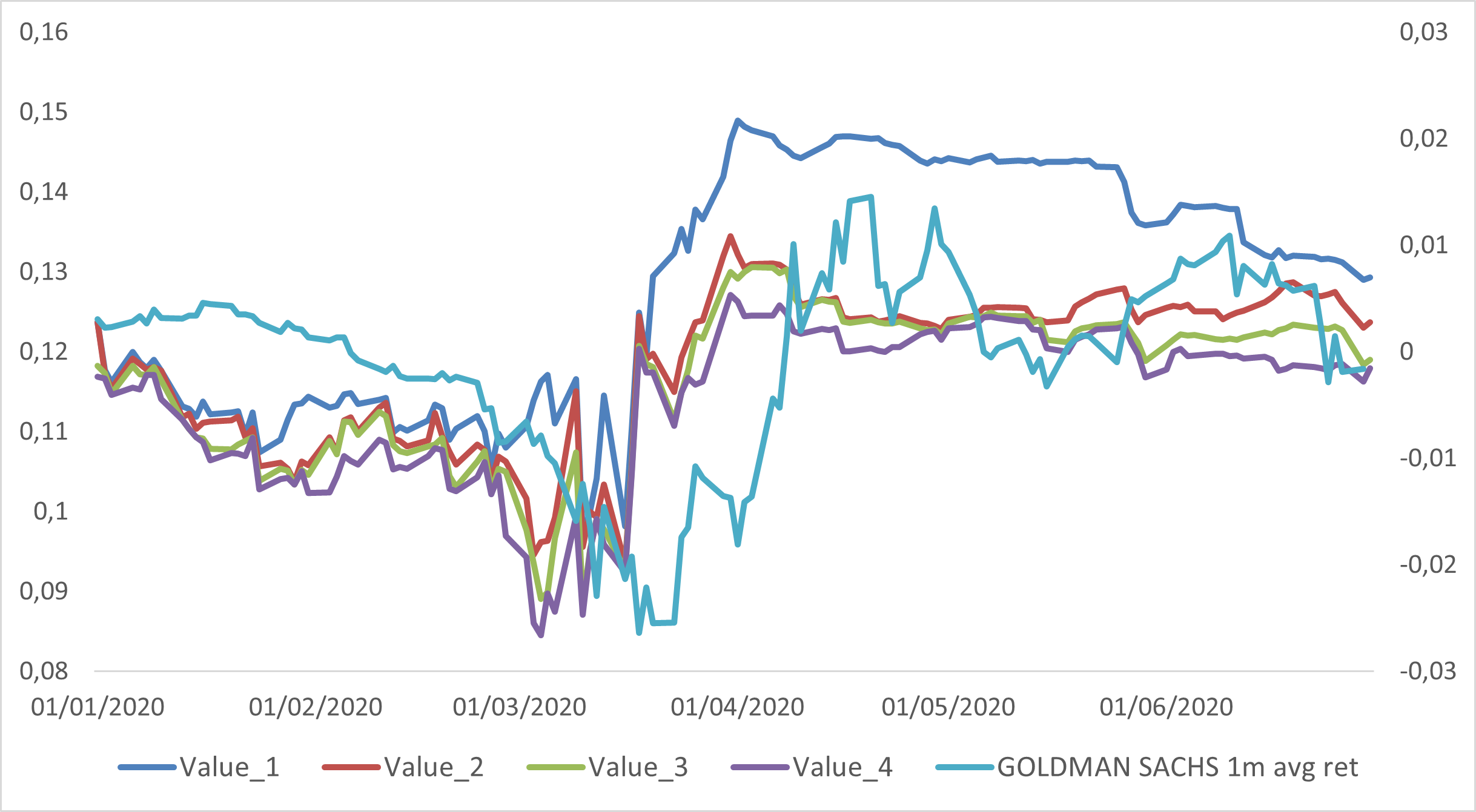}
\caption{Versus GS 1m average returns}
\end{subfigure}\hfill
\begin{subfigure}{0.45\textwidth}
\includegraphics[width=\textwidth]{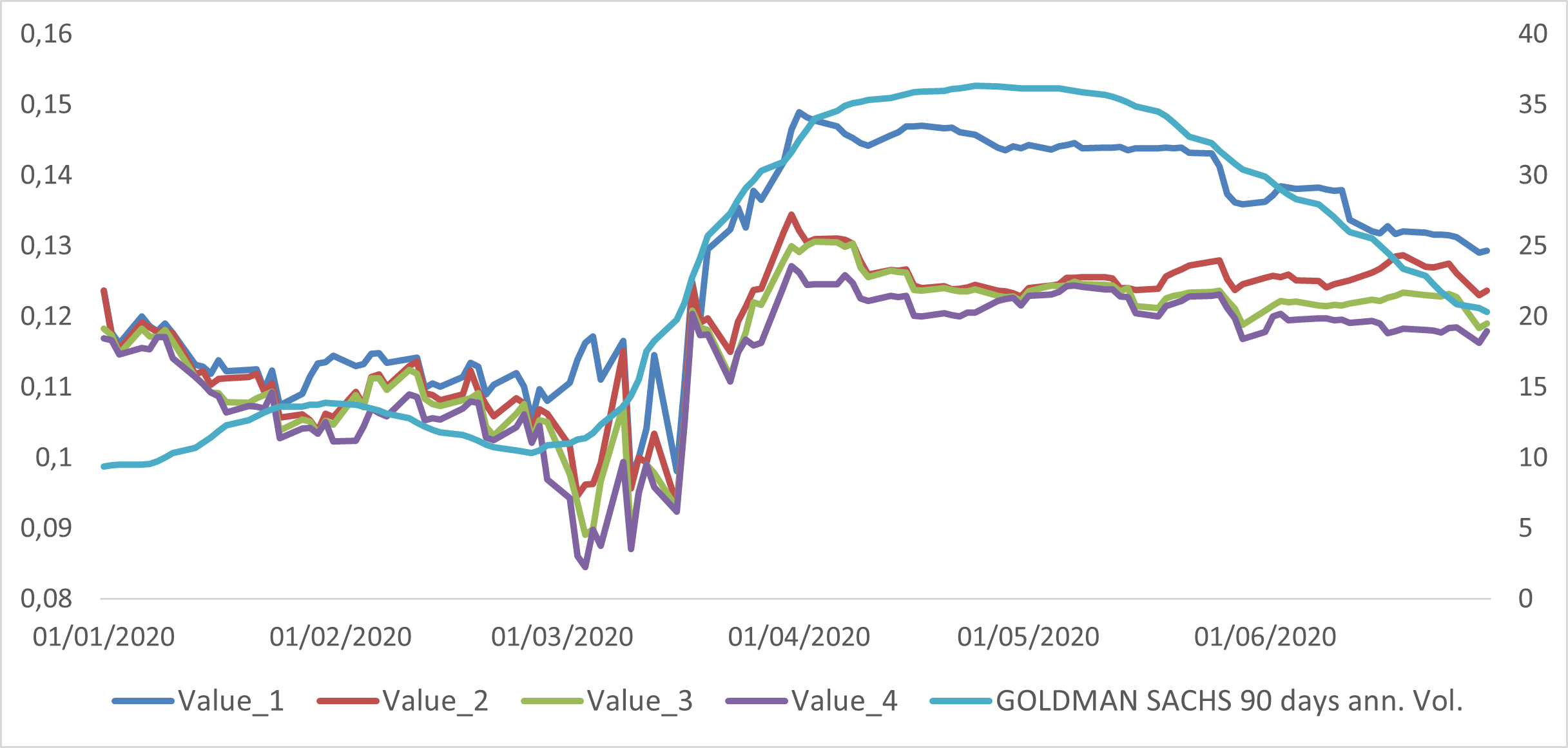}
\caption{Versus GS 90-days realized volatility}
\label{histdist}
\end{subfigure}\par
\vskip\floatsep% normal separation between figures
\begin{subfigure}{0.45\textwidth}
\includegraphics[width=\textwidth]{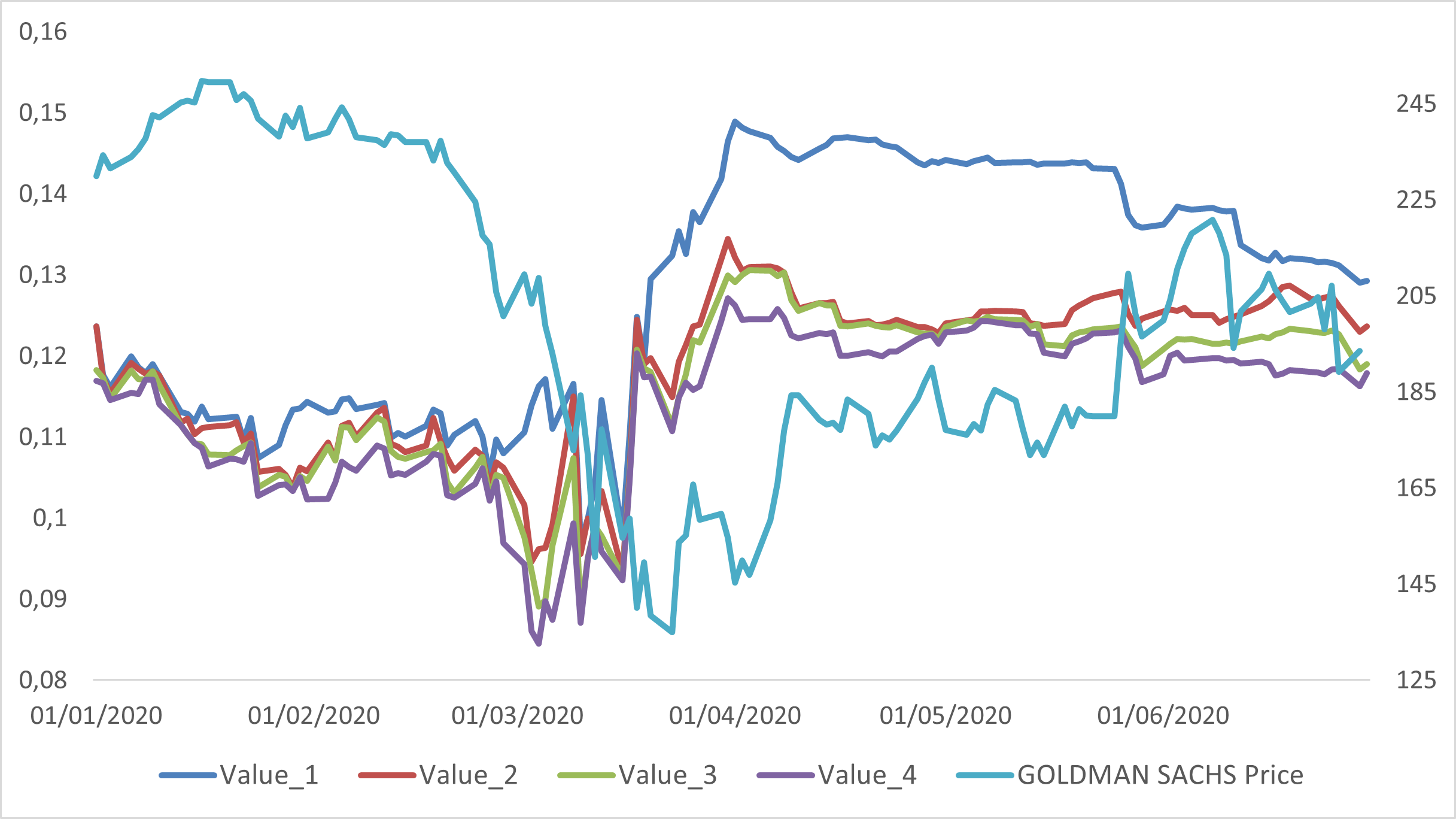}
\caption{Versus GS stock price}
\label{wishart}
\end{subfigure}
\medskip
\medskip
\medskip
\captionsetup{skip=10pt}
\caption{This figure represents the time series dynamics of the indicator $\sigma_{\Gamma}$ concerning the top four driver candidates impacting the performance of Goldman Sachs (GS) stock during the COVID-19 crisis. These candidates are identified based on the highest cumulative sum values for the indicators among a pool of 109 candidates. The figures presented in this analysis depict the time series behavior of the indicators for the identified top causal candidates, denoted as values 1 to 4.}
\label{figureretvolprice}
\end{figure}

\begin{figure}[h!]% [H] is so declass\'e!
\centering
\begin{subfigure}{0.45\textwidth}
\includegraphics[width=\textwidth]{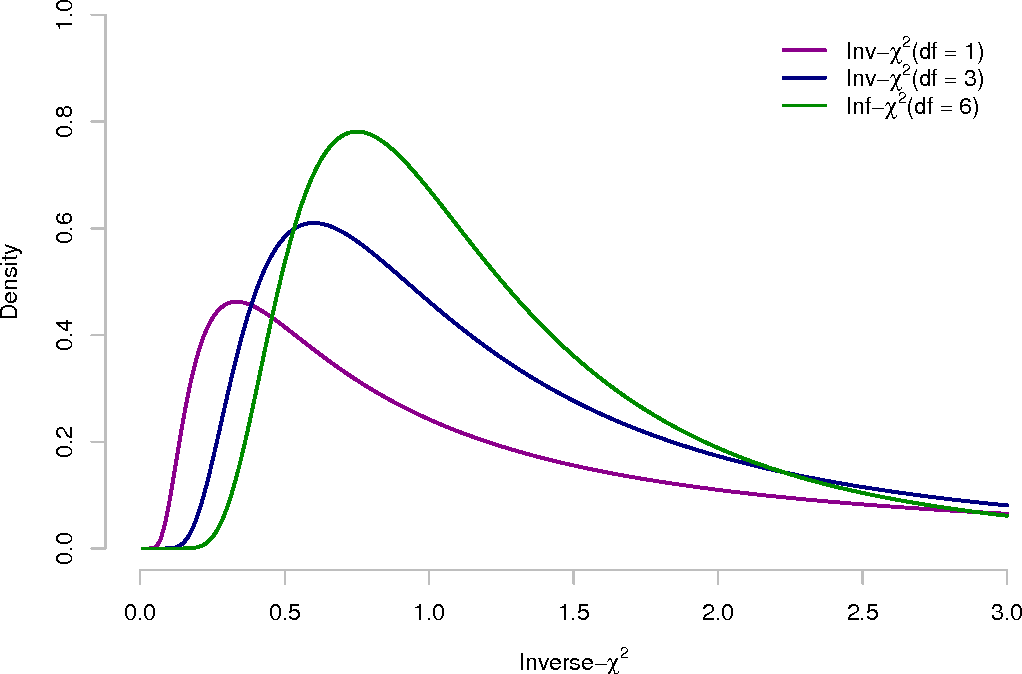}
\caption{Theoretical inverse Wishart distribution from \citep{Nydick2012TheWA}.}
\label{wishart}
\end{subfigure}\hfill
\begin{subfigure}{0.45\textwidth}
\includegraphics[width=\textwidth]{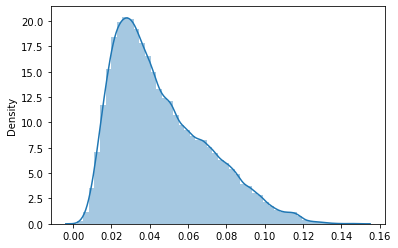}
\caption{All drivers' $\sigma_{\Gamma}$ historical distribution for Goldman Sachs stock from 2014 to 2023.}
\label{histdist}
\end{subfigure}\par

\begin{subfigure}{0.60\textwidth}
\includegraphics[width=\textwidth]{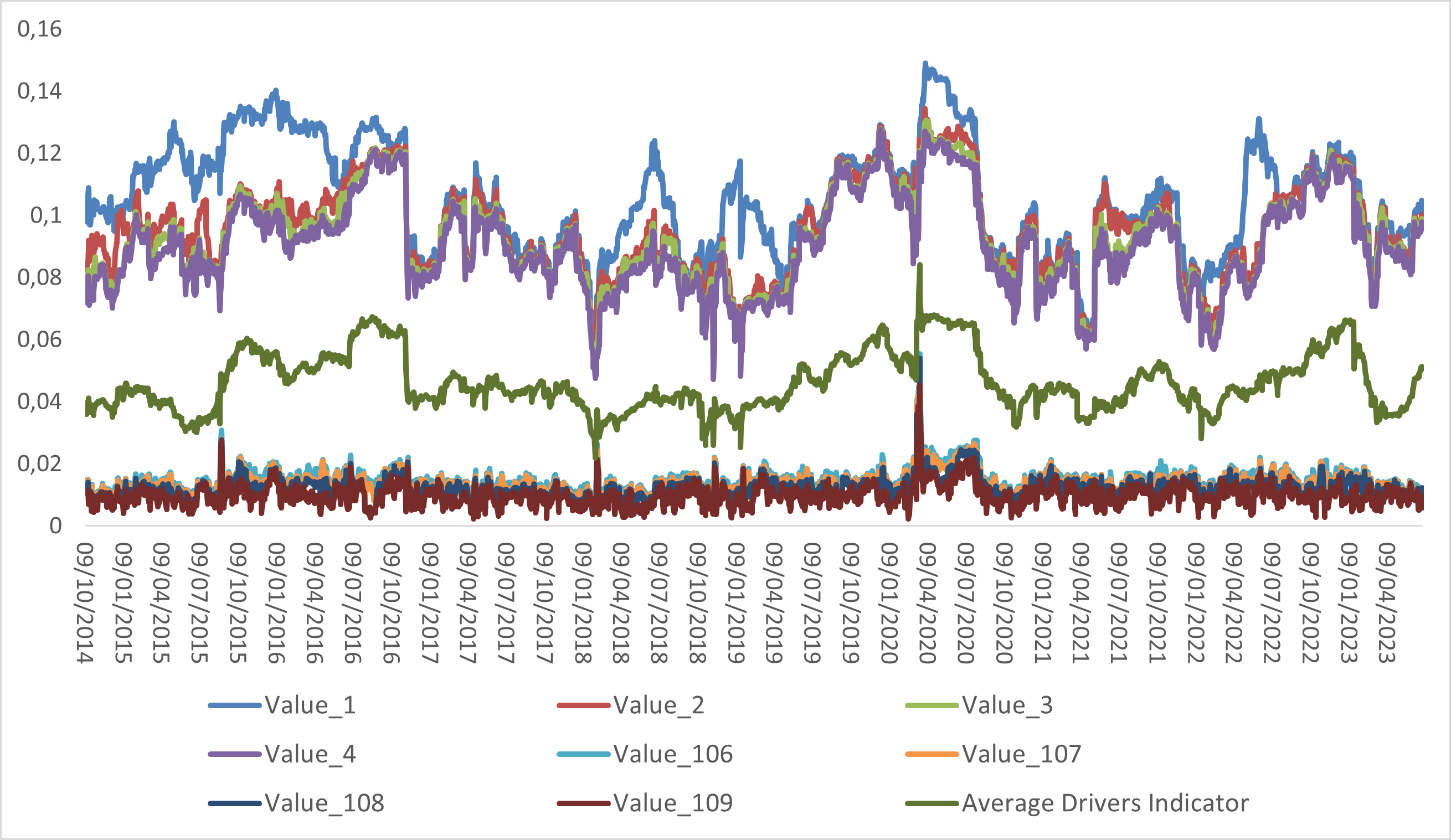}
\caption{time series of the indicator $\sigma_{\Gamma}$ values for some of the top, bottom, and average values' cases, for Goldman Sachs stock from 2014 to 2023 as the effect variable.}
\end{subfigure}
\medskip
\medskip
\medskip
\caption{$\sigma_{\Gamma}$ historical distribution values for Goldman Sachs (GS) as effect variable between 2014 and 2023 and comparison with theoretical Inverse Wishart distribution}
\label{CausalDistr}
\end{figure}

In Figure \ref{figureretvolprice}, the top four causal drivers candidates' $\sigma_{\Gamma}$ values for Goldman Sachs during the peak of the COVID crisis are shown. The ranking changes for the causal variables candidates during the period, but the top four values are recorded and shown in the plots. These indicators' values are compared with the Goldman Sachs stock price, volatility, and average returns time series. In Figure \ref{figureretvolprice}, the predictability of the 1-month average return and realized volatility from the top indicators for the causal drivers \textls[-25]{candidates' sample, can be seen. Also, a strong inverse relationship between these indicators and the stock price occurs.} All of these experiments show that the indicators with a causal driver having the highest values are strong risk analysis measures indicators, which include prediction (causal) capabilities for the stock. In Figure \ref{CausalDistr}, the distribution of all causal drivers' $\sigma_{\Gamma}$ candidates' values is analyzed for the cases in which Goldman Sachs' stock time series returns are the effect variable from 2014 to 2023. The historical distribution in Figure \ref{histdist} follows an inverse Wishart \textls[-20]{distribution, which can be contrasted with Figure \ref{wishart}.\par }
At this stage, the capabilities of the indicator in detecting structural changes has been verified, with the ability to monitor them at each timestamp, instead of using estimation windows such as in the Granger test, the Chow test, or other likelihood ratio tests. It has shown strong predictability capabilities with respect to the effect variable, with the method itself not based on predictability between the cause-and-effect variable, as in the Granger test. It instead focuses on the distribution properties of the explanatory power of the spectra from lagged correlation matrices. Therefore, the predictability is a consequence and not an assumption of the method, which makes this predictability closer to true causal interactions. To confirm the hypothesis of causal interactions, another experiment is shown in which the causal relationship is known in advance and tested with the indicator.\par

\begin{figure}[h!]

\centering

\begin{minipage}{\textwidth}
    \centering
    \includegraphics[width=130mm]{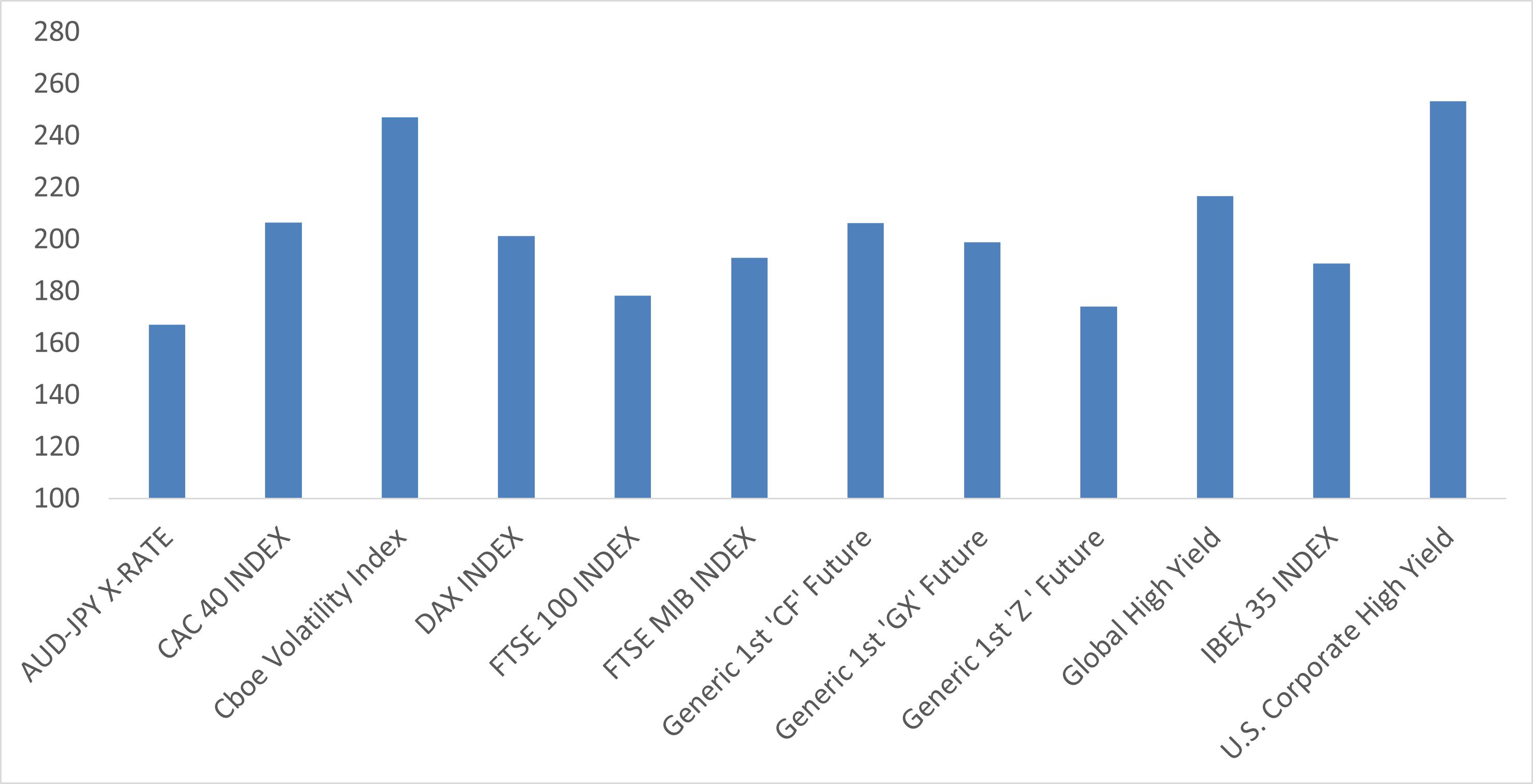}
    \captionsetup{skip=10pt} 
    \caption{The top twelve causal drivers' $\sigma_{\Gamma}$ candidates daily time series sum values for the U.S. credit default swap high yield index as the effect variable.}
    \label{CDX_full}
\end{minipage}

\vspace{\floatsep}

\begin{minipage}{\textwidth}
    \centering
    \includegraphics[width=130mm]{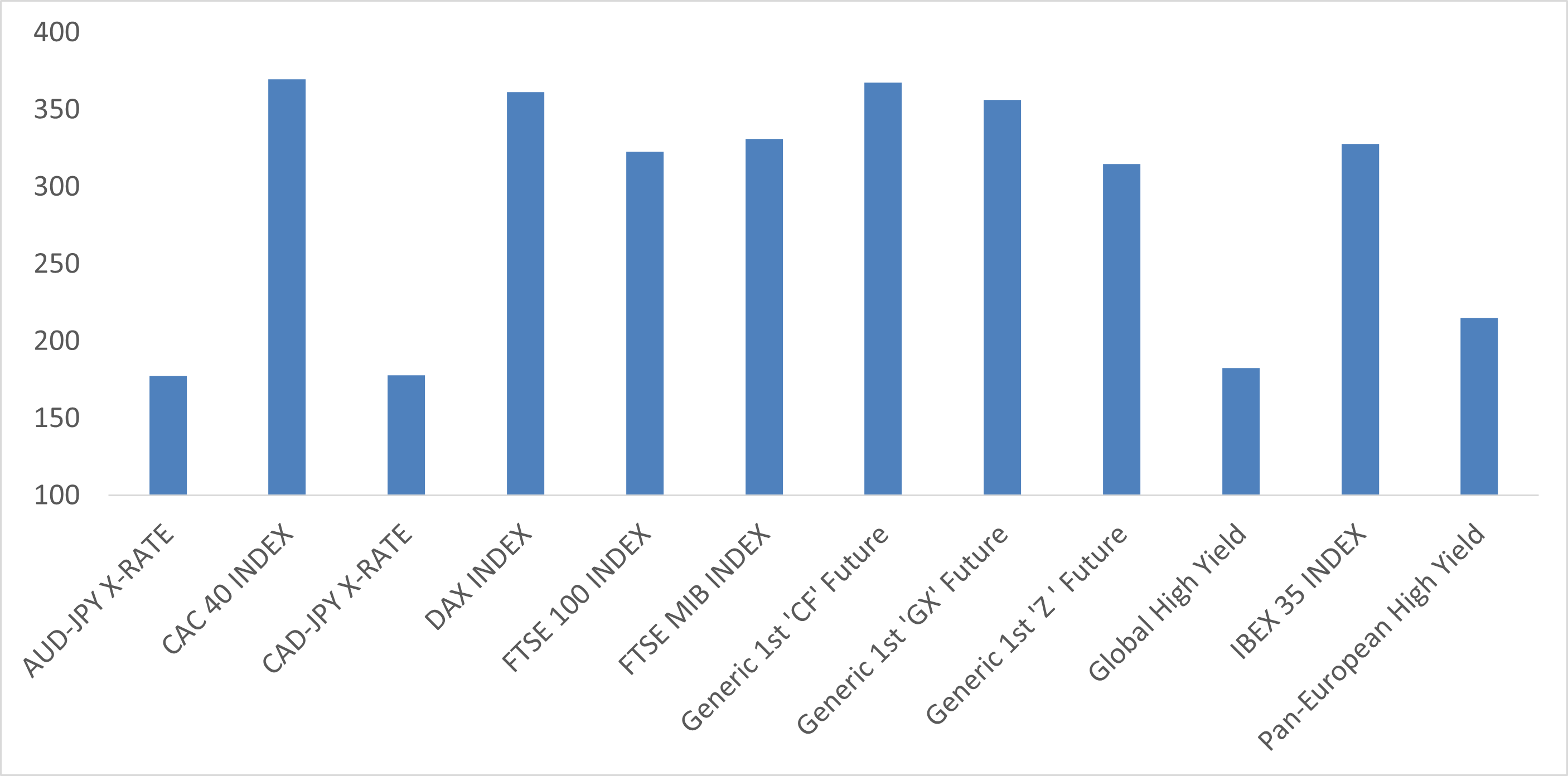}
    \captionsetup{skip=10pt} 
    \caption{The top twelve causal drivers' $\sigma_{\Gamma}$ candidates daily time series sum values for the Stoxx 600 equity index as the effect variable.}
    \label{stoxx600_full}
\end{minipage}
\end{figure}

Experiments are conducted consisting of selecting a known causal relationship in the financial markets, and checking if the rule for the maximum $\sigma_{\Gamma}$ indicator is able to select the same pair of cause-and-effect variables, given a particular effect variable, and from more than a hundred causal drivers' candidates. For the effect variable, the credit default swap (CDS) index for high yield US corporates (CDX), the Stoxx 600 equity index, the Stoxx volatility index, and the CDS investment grade for the Europe corporate sector (iTraxx EU) were used. The indicator is given in the form of a daily time series, and in order to measure the level of causal interactions, all of these values are added up and displayed in a bar graph. Figure \ref{CDX_full} shows how the cumulative sum of all of the indicator's values during the period from 2014 to 2023 for the CDX index as effect candidate identifies the US high yield corporate credit as the top causal driver candidate couple. This is the one among all of the 110 candidates which has the highest prior probability to be a causal variable for the CDX. The reason is that the insurance contract on the high yield US corporates credit is the CDX index, so in consequence it is causally connected with the dynamics of that type of credit itself. 

\begin{figure}[h!]
\centering

\begin{minipage}{\textwidth}
    \centering
    \includegraphics[width=130mm]{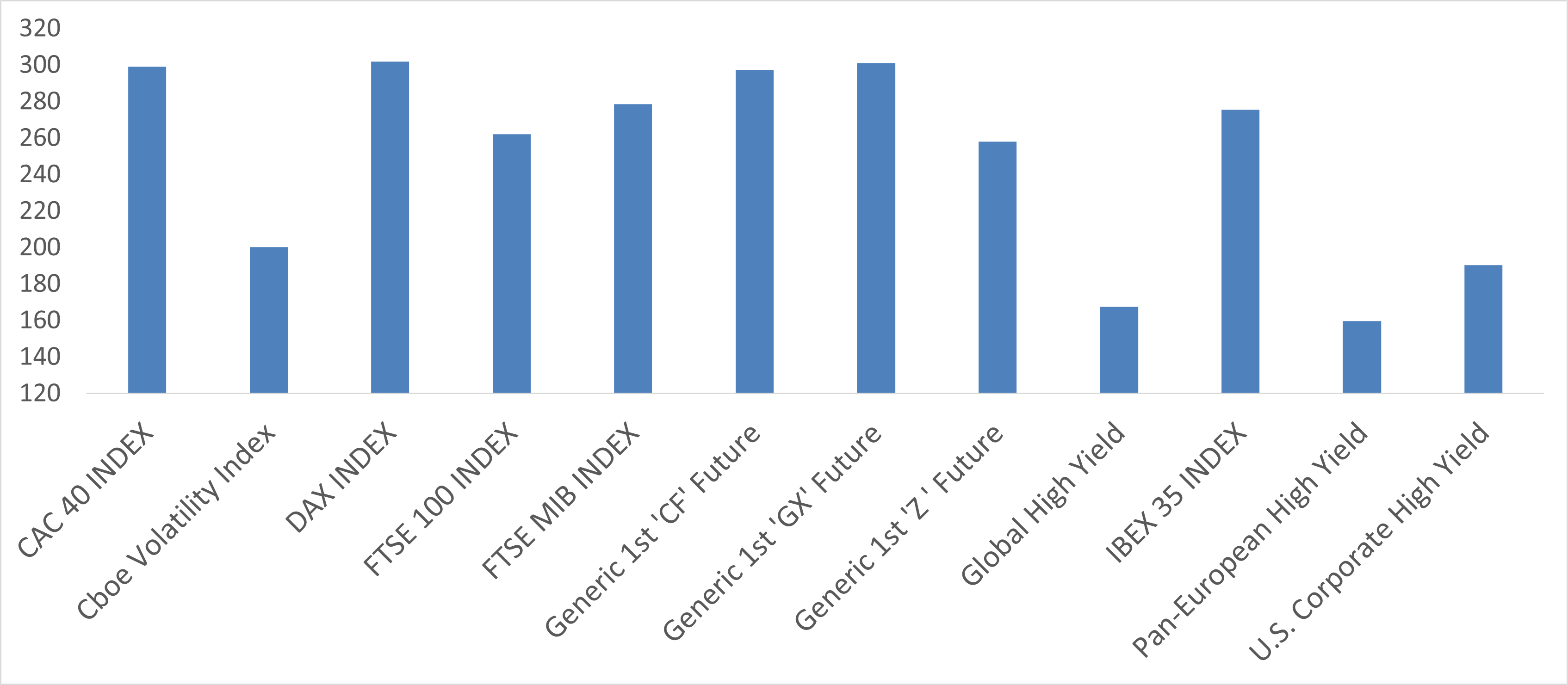}
    \captionsetup{skip=10pt} 
    \caption{The top twelve causal drivers' $\sigma_{\Gamma}$ candidates daily time series sum values for the Stoxx 600 volatility index as the effect variable.}
    \label{VSTOXX_full}
\end{minipage}

\vspace{\floatsep}

\begin{minipage}{\textwidth}
    \centering
    \includegraphics[width=130mm]{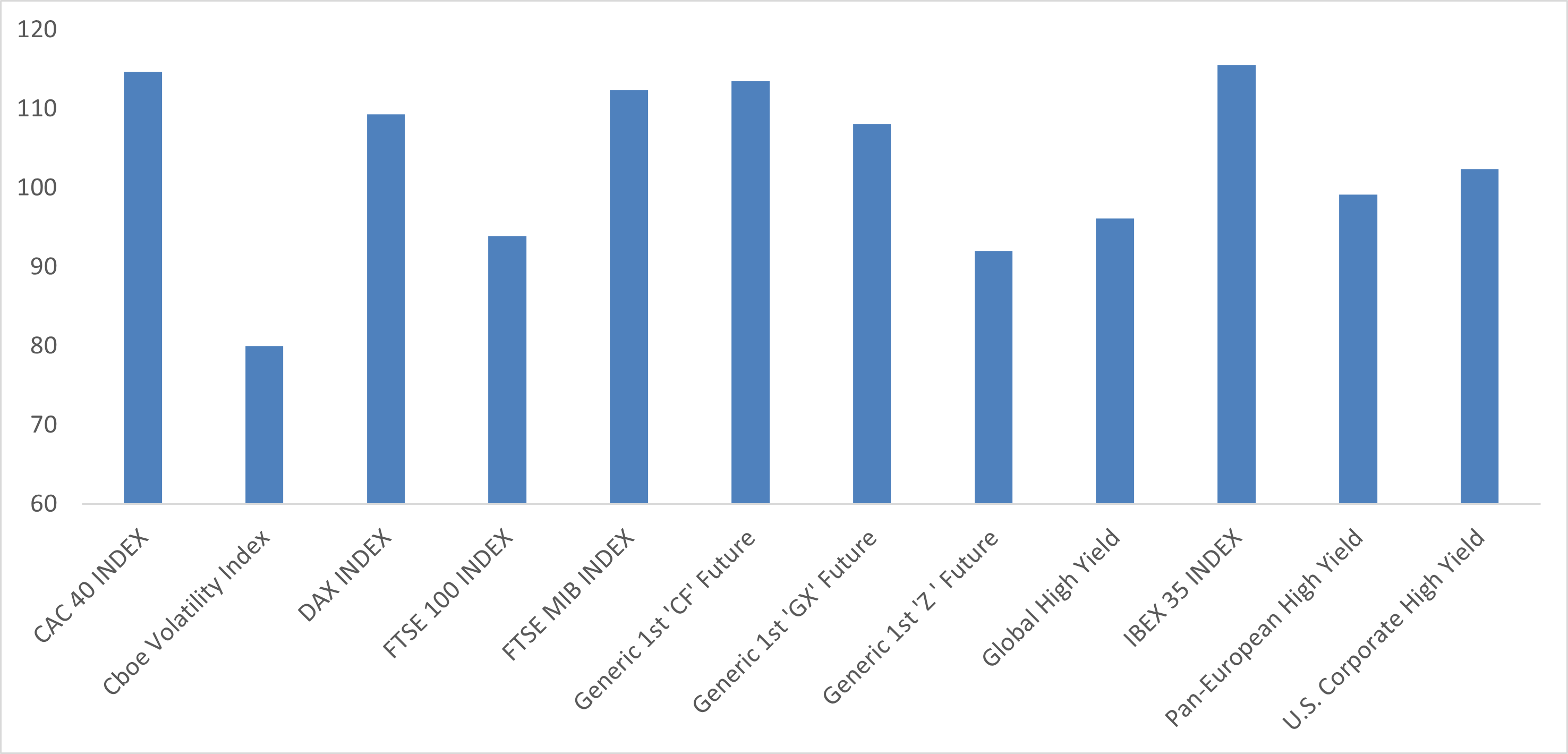}
    \captionsetup{skip=10pt} 
    \caption{The top twelve causal drivers' $\sigma_{\Gamma}$ candidates daily time series sum values for the iTraxx (Europe) credit default swap investment-grade index as the effect variable.}
    \label{ITRAXX_EU_Full}
\end{minipage}

\end{figure}

Next, Figure \ref{stoxx600_full} shows, for the Stoxx 600 index, which consists of the largest companies' stocks in Europe by market capitalization, the strongest indicators' cumulative sum values for the main equity indexes of the European countries. This is obvious, as both sets of indexes share the same constituents. In the case of the Stoxx volatility index, the same rationale applies as for the Stoxx 600 index (Figure \ref{VSTOXX_full}). Being an implied volatility index, the constituents are not shared this time by both sets of indexes, however the stocks and their indexes must be causally connected with their implied volatility, and as a consequence, the one in the index. In the case of the iTraxx Europe CDS index, displayed in Figure \ref{ITRAXX_EU_Full}, the main European equity indexes are shown as the causal counterparts. Not having a European credit investment grade index in the sample demands the choice of using the equity component of these corporates instead. This equity component has strong connections with the credit component and their CDS index, and from the sample, are known to be the most causally connected, as they are the stock prices of the corporate names in which the default insurances are written (all of the CDS contracts that make up the iTraxx index).

\begin{figure}[h!]
\setlength\abovecaptionskip{0\baselineskip}
	\centering
	\includegraphics[width=130mm]{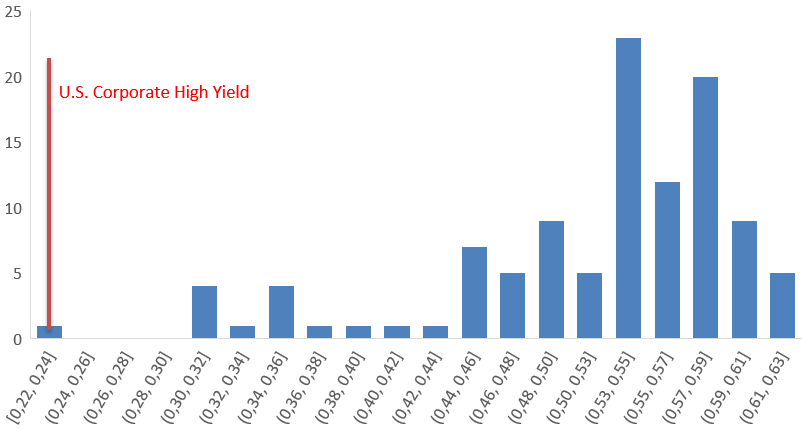}
 \medskip
	\caption{Distribution of the average values of the Jensen-Shannon distance between the historical distribution of the indicator $\sigma_{\Gamma}$ and the inverse Wishart distribution fitted to this historical data, using 1, 3, and 6 degrees of freedom for the latter, and averaging the three distances (for all cases in which the CDX index is the effect and the full sample of causal candidates is the causes).}
	\label{jensenshannon}
\end{figure}

In Figure \ref{jensenshannon}, the distributions of the average values of the Jensen-Shannon distance are shown, between the historical distribution of the indicator $\sigma_{\Gamma}$ and the inverse Wishart distribution, for 1, 3, and 6 degrees of freedom, fitted to the historical data of the indicators. Values of the three distances for the three degrees of freedom are averaged. The distribution of the average is shown for the cases in which the effect variable is the CDX index and the causes variables consist of all of the candidates from the sample of more than one hundred candidates. It can be seen in Figure \ref{jensenshannon} that the top causal variable candidate for the CDX index, from experiments and from common knowledge, is the U.S Corporate high yield, as previously stated in the paper, which has the indicator with the shortest distance between the historical distribution and the fitted inverse Wishart for different degrees of freedom. This experiment verifies the hypothesis that stronger causal interactions measured by the indicator make the indicator behave more like an inverse Wishart distribution.

\subsection{Causal liquidity risk indicator for brokerage/banking firms}
For the brokerage dataset, the daily brokerage activities, with time series data of 15,000 clients approximately, is analyzed. The goal is to investigate the power of the structural temporal (causal) interactions between the cash-balance of the company and other related variables from the business as a function of time. Experiments are performed with cause-and-effect pairs of variables, with the effect being the cash-balance. For illustration, the results are plotted in thematic buckets, such that only the cause variables are displayed and the effect variable is understood to be the same in all cases.\par 
The buckets consist of: 
\begin{itemize}
    \item The account status, which includes the daily amount in account currency of open positions in different products, including shares, bonds, mutual funds, CFDs, derivatives, FX Spot, and ETO.
    \item Bookkeeping cash, which includes all of the daily cash movements in the account currency, both internal and external cash transactions, and the financial product which is the reason for the transaction.
    \item Trades executed, which is focused on the daily buying-and-selling transaction activities in the account currency for all clients and in all broker products.
\end{itemize}
A description of the causal candidates' variables is shown in Table \ref{brokeragevariables}. The cash-balance of the company is the effect variable for all cases and figures, and the plots display all indicators' time series with their respective causal counterpart as a label.

\begin{table}[h!]
\begin{center}
\caption{Brokerage business variables}
\vspace{5mm}
\resizebox{\textwidth}{!}{%
\begin{tabular}{ll}\hline
Variable                                 & Description                                                                \\
\hline
Bonds                                    & Fixed income products                                                      \\
CFD                                      & Contract for Difference (Derivative that replicates a   financial product) \\
ETO                                      & Exchange Traded Options                                                    \\
Futures                                  & Future contract                                                            \\
FX (Spot and Forward)                    & Currency exchange rates                                                    \\
FX option                                & Options on exchange rates                                                  \\
Shares                                   & Companies' stocks                                                          \\
Mutual Fund                              & Third-party managed funds                                                  \\
CFD open\ positions                 & Managed Volumes of CFD                                                \\
ETO open\ positions             & Managed Volumes of ETO                                                   \\
Futures open\ positions         & Managed Volumes of Futures                                               \\
FX open\ positions              & Managed Volumes of FX                                                   \\
Mutual Funds open\ positions & Managed Volumes of Mutual Funds                                           \\
Stocks open\ positions           & Managed Volumes of Stocks                                                 \\
Bonds open\ positions          & Managed Volumes of Bonds                                           \\
Bonds\_BKCash                            & Cash flows from/for Bonds positions                                         \\
CFDs\_BKCash                             & Cash flows from/for CFDs positions                                          \\
Contract Futures BKCash                  & Cash flows from/for Futures positions                                       \\
Contract Options BKCash                  & Cash flows from/for ETO    positions                                        \\
FX Option BKCash                         & Cash flows from/for FX options positions                                    \\
Spot \& Forward BK Cash                  & Cash flows from/for FX positions                                            \\
mutual\_funds BK Cash                    & Cash flows from/for Mutual Funds positions                                  \\
Shares BKCash                             & Cash flows from/for Shares positions \\\hline             \label{brokeragevariables}                      
\end{tabular}}
\end{center}

\end{table}

\begin{figure}[h!]
\centering

\begin{minipage}{\textwidth}
    \centering
    \captionsetup{skip=10pt} 
    \includegraphics[width=150mm]{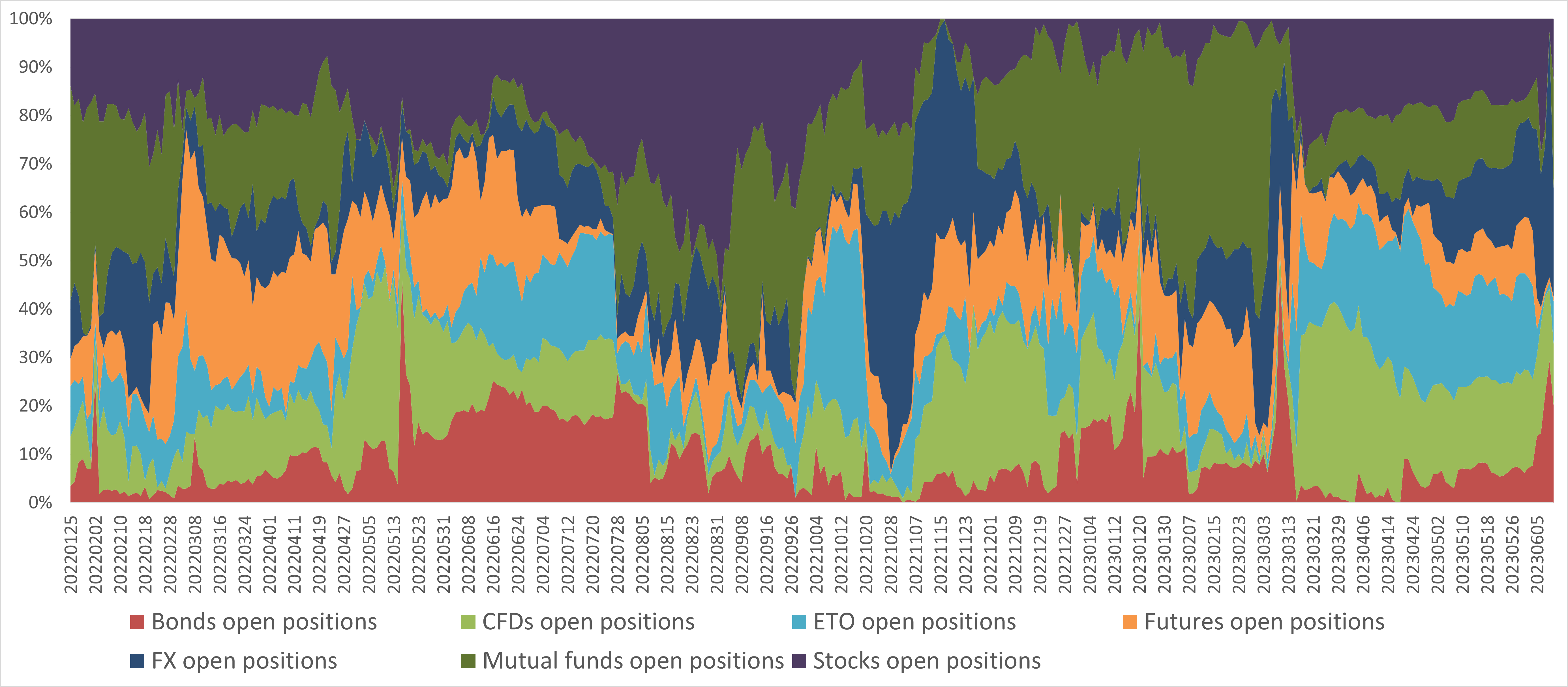}
    \caption{Account status, value of $\sigma_{\Gamma}$ for $\max$ lag = 2, cash-balance vs. broker securities open positions.}
    \label{figureL2W60Acc}
\end{minipage}

\vspace{\floatsep}

\begin{minipage}{\textwidth}
    \centering
    \captionsetup{skip=10pt} 
    \includegraphics[width=150mm]{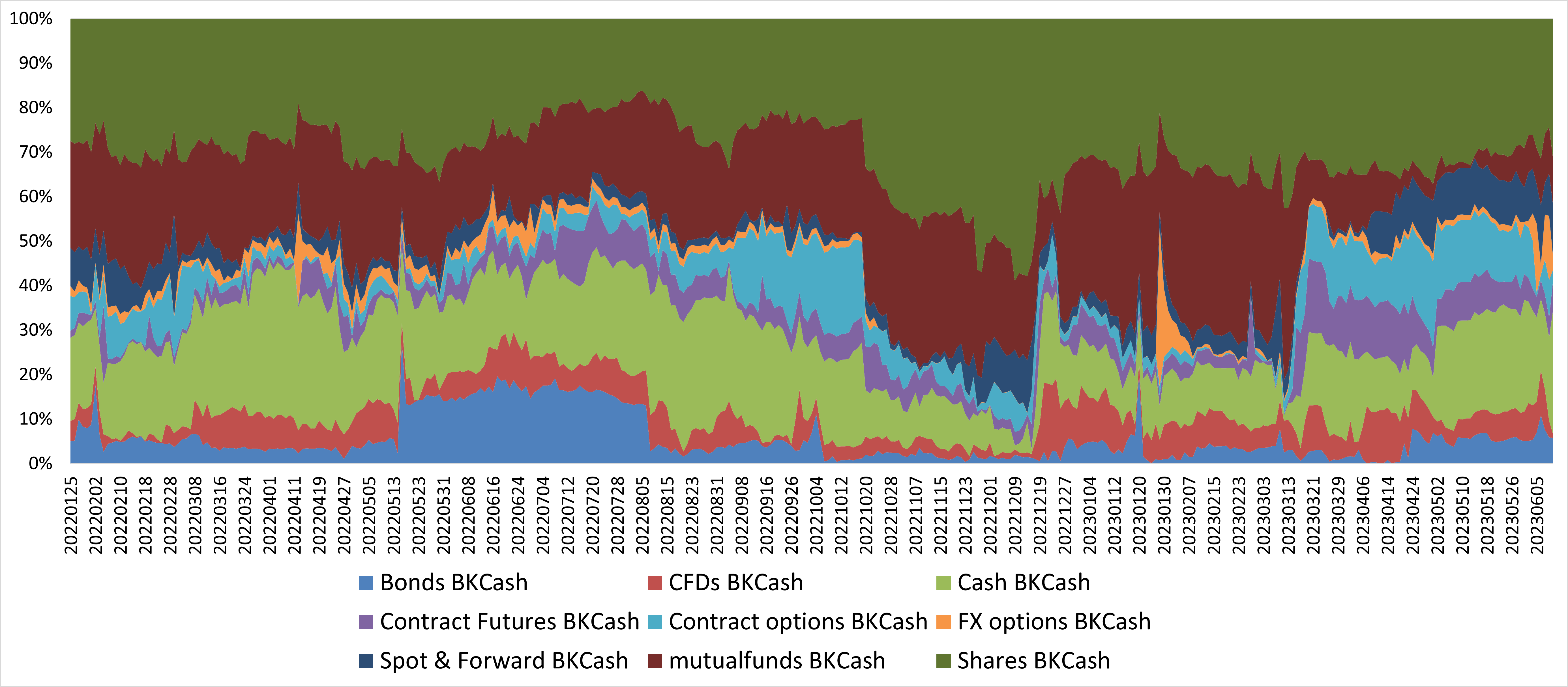}
    \caption{Bookkeeping cash, value of $\sigma_{\Gamma}$ for $\max$ lag = 2, cash-balance vs. broker cash transactions.}
    \label{figureL2W60B}
\end{minipage}

\end{figure}

\begin{figure}[h!]
\setlength\abovecaptionskip{0\baselineskip}
	\centering
	\includegraphics[width=150mm]{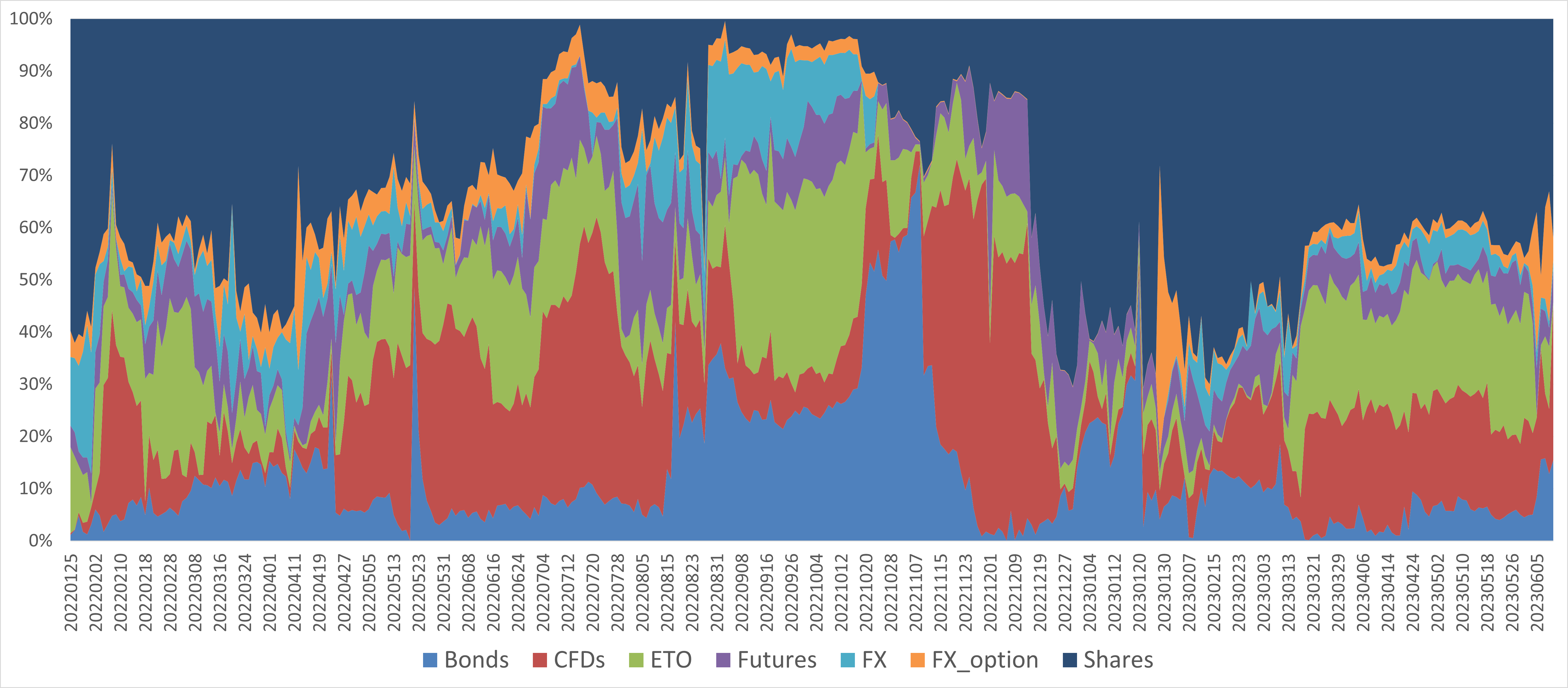}
	\caption{Broker trades executed vs. cash-balance, value of $\sigma_{\Gamma}$ for $\max$ lag = 2.}
	\label{figureL2W60TE}
\end{figure}

Figures \ref{figureL2W60Acc}---\ref{figureL2W60TE} show area plots in which the values of the indicator for each causal candidate variable are compared, so that all values for the indicators amount to $100\%$ for each date, and the time series is displayed with higher areas indicating causal interactions between them and the cash-balance. From the experiments, it can be seen that the account status bucket has more mixed results, with mutual funds from November 2022 until March 2023 comprising the majority. After that, the majority of cash movements can be explained by stocks (shares), as can be seen in Figure \ref{figureL2W60Acc}.

In the Bookkeeping cash bucket, shown in Figure \ref{figureL2W60B}, a sustained pattern of the majority of cash transactions coming from shares activities is presented. Until March 2023, this majority was shared with mutual funds, however this ceased from March onwards, and the derivatives and FX products have started to build importance since then. The relationship with daily variations of cash-balance gross amount (net of outflows and inflows for each closing day) is studied,  with respect to the causal candidates' variables, each in terms of net movements of cash for a closing day (all outflows plus inflows).

Finally, the Trades executed bucket shows very similar behavior to the Bookkeeping cash bucket, which serves as a validation point for our results, as can be seen in Figure \ref{figureL2W60TE}. The reason why the mutual funds product is not in this figure is because its execution process is different, as much of the operations do not need to be converted to cash (funds transfer). This could mean that, although mutual funds has been a product that, at least historically until March 2023, was responsible for the movement of cash-balance from the clients (seen in the Bookkeeping bucket in Figure \ref{figureL2W60B}), this was not due to daily executions in mutual funds with cash, but in the form of internal and external transfer funds. With regards to Trades executed, we can see that from June 2022 until December 2022, the relationships were mixed, with shares not holding the majority. From December 2022 until now, shares have held the majority with some derivatives gaining importance since March 2023.

\begin{figure}[h!]
\centering

\begin{minipage}{\textwidth}
    \centering
    \captionsetup{skip=10pt} 
    \includegraphics[width=150mm]{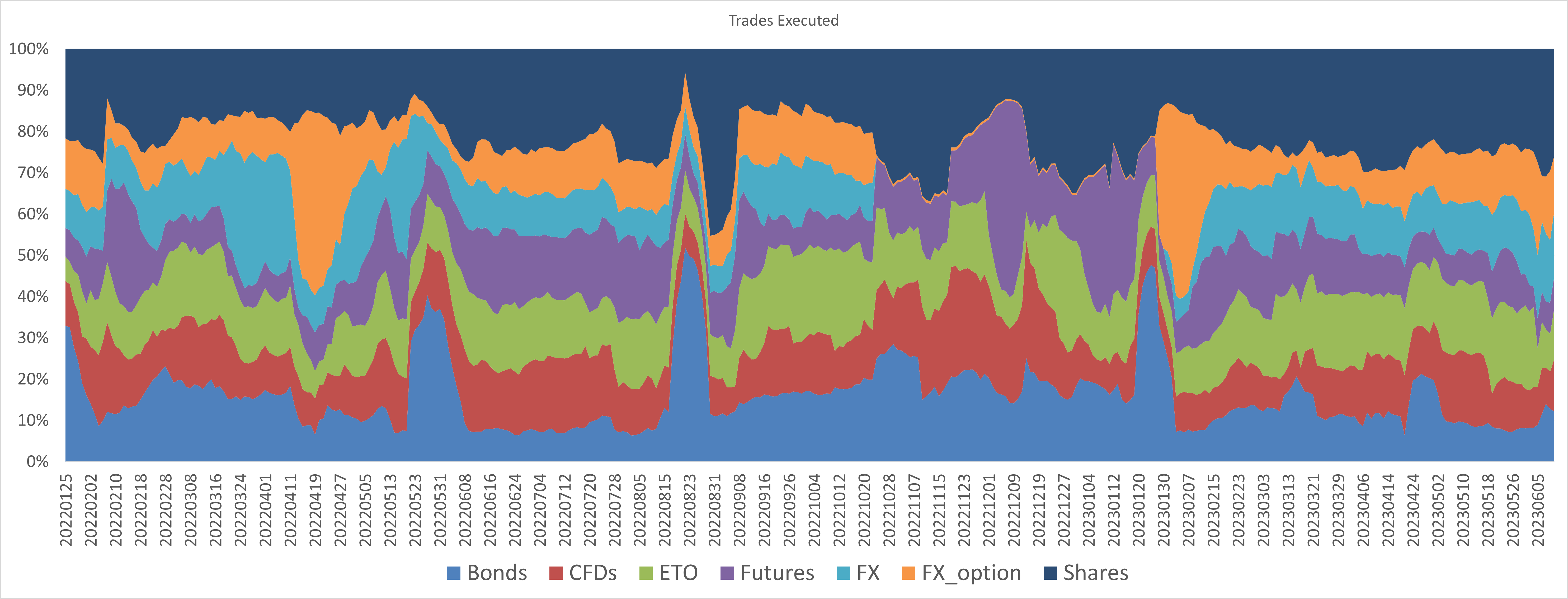}
    \caption{Broker trades executed vs. cash-balance, value of $\sigma_{\Gamma}$ for $\max$ lag = 10.}
    \label{figureL10W60TE}
\end{minipage}

\vspace{\floatsep}

\begin{minipage}{\textwidth}
    \centering
    \captionsetup{skip=10pt} 
    \includegraphics[width=150mm]{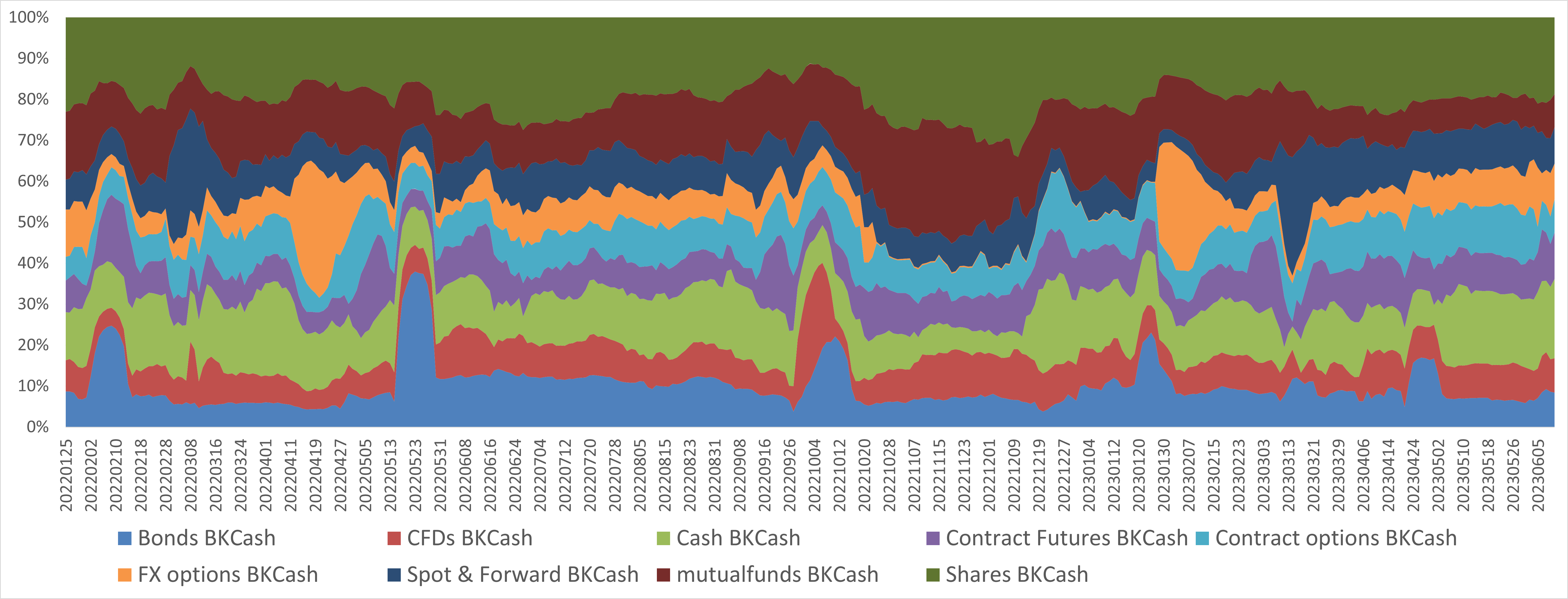}
    \caption{Bookkeeping cash, value of $\sigma_{\Gamma}$ for $\max$ lag = 10, cash-balance vs. broker cash transactions.}
    \label{figureL10W60B}
\end{minipage}

\end{figure}

With greater lags (five or more) values only get smoother but patterns look similar (in Figures \ref{figureL10W60TE} and \ref{figureL10W60B}, results for 10 lags are displayed). This verifies the theoretical setup presented in this document in that, for greater lags, the standard deviation of their largest eigenvalue varies more. This implies a higher probability that the two variables deviate from independence toward a structured (causal) relationship. This is reflected by the fact that the indicator becomes less volatile in time.

All of the findings from the Figures \ref{figureL2W60Acc}---\ref{figureL2W60TE} are aligned with the rationale and causal behavior of the cash-flows, given the brokerage business of the dataset. The majority of the client operations were made on stocks, so it makes sense that their executions and cash transactions have more causal interactions with the cash-balance of the company (Figures \ref{figureL2W60Acc} and  \ref{figureL2W60B}). Periods of weak shares business operations (executions) are compensated by the second and third most traded products, the bonds and CFDs. This is true for executable products, in the case of mutual funds where the relationship is only via cash transaction or fund transfer (fund transfers do not affect the cash-balance so they are omitted). The fact that the Account status bucket gives mixed results in Figure \ref{figureL2W60Acc} can be explained because this bucket consists of balances (levels), and increments in cash-balance are not related by causal interactions with other balances' levels. In contrast, increments of cash-balance must be related to other cash-flows within the system by causal interactions, which is well-captured in the other two buckets (cash transactions and executions). Useful findings from the experiments for brokerage business for retail and institutional clients (ie., banks) include:

\begin{itemize}
    \item Liquidity risk can be monitored by analyzing interactions between operational flows (executions and trading activities) and increments in the cash-balance.
    \item As most brokerage businesses have stocks as the main traded asset, their flows should be monitored first and are highly correlated to the equities market performance (crisis and rallies would have an impact on liquidity risk).
    \item Mutual funds do not have an impact on liquidity as most operations are being made via fund transfers, which do not convert the asset to cash.
    
\end{itemize}

Finally, with regard to the limitations of the indicator, it can be said that it works better with data of higher frequencies than daily, as it can deal with a higher number of lags, in contrast to daily time series, in which more than ten lags is redundant. The fact that it only works for pairs of time series is limiting for cases in which more simultaneous relationships need to be detected.

%\begin{table}[H]
%\begin{center}
%\caption{Caption of the table. Tables should be placed in the main text near to the first time they are cited.}
%\begin{tabular}{ccc} \hline
% & & \\\hline
% & & \\
% & & \\
% & & \\
% & & \\
% & & \\
% & & \\
% & & \\
% & & \\
% & & \\\hline
% & Note:(Table body should be created by MS word table function; three-line table is preferred.)
%\end{tabular}
%\end{center}
%\end{table}

%\begin{figure}[H]
%\begin{center}
%\includegraphics{figure.pdf}
%\caption{Legend of the figure. If there are multiple panels, they should be listed as: (a) Description of what is contained in the first panel; (b) Description of what is contained in the second panel. Figures should be placed in the main text near to the first time they are cited. A caption on a single line should be centered. (All figures should be in the RGB color mode, and be provided as separate files. Image resolution should be a minimum of 300 dpi.)}
%\label{Fig1}
%\end{center}
%\end{figure}

\section{Conclusions}

A method to measure causal interactions between pairs of time series using the variability in explanatory power of the largest eigenvalue is presented. The method is only applied and can deliver the expected results with no more than a pair of time series, one cause and one effect candidate. Using the homology of the interactions of two walls pushing and pulling a Coulomb gas, which can be modelled by the largest eigenvalue of a random matrix following a Tracy-Widom distribution, the standard deviation of the explanatory power of the largest eigenvalue is used to measure a similar interaction in a system consisting of two time series. The causality component of the interaction is enforced by the direction of time, thanks to lagging the correlation matrices from which the variations of the explanatory power are being measured. To the authors' knowledge, it is the first time random matrix theory and eigenvalues distributions have been applied to measure causal interactions, and also that an experimental system in statistical physics has been used as homology for this purpose. Experiments were conducted with synthetic and real time series and scenarios, including known causal relationships. The method can capture these interactions more accurately than other traditional methods such as the Granger causality test. The indicator can detect structural breaks in time series as shown by experiments with different ARMA models. It can also detect widely known causal variables' candidates for a given set of effect variables among more than one hundred causal candidates (for example, the CDS vs credit as in Figure \ref{CDX_full}). Two datasets are used for market and liquidity risk analysis based on the indicator. The simplicity, dynamism, and the fact that it does not rely on predictability like other methods, but on spectral theory, random matrix theory, and the homology with statistical and experimental physics properties of causality, make the method extremely useful for risk management with time series for financial institutions and regulators.\par

For future work, it would be interesting to investigate indicators that can detect causal interactions among more than two time series. The use of random matrix theory seems promising in adding structure to the problem of causal interactions detection. It would be interesting to study the connection between the inverse Wishart distribution, which the causal interactions indicator follows, and the Coulomb gas analogy. It would also be interesting to study configurations of the Coulomb gas analogy for the case of multiple connected gas lines (as in a causal graph of interactions). Finally, it would be interesting to test the indicator as a source of alpha in trading strategies in which information from options derivatives is used to predict their underlying stock price future movements. Also, in the study of prepayments behavior of loans or mortgages, it can be used to detect which variables cause these prepayments, with strong implications in household economies.

%%%%%%%%%%%%%%%%%%%%%%%%%%%%%%%%%%%%%%%%%%%%%%%%%%%%%%
%          AI TOOLS, USE AND LOCATION
%%%%%%%%%%%%%%%%%%%%%%%%%%%%%%%%%%%%%%%%%%%%%%%%%%%%%%
%We follow COPE's guidelines and policies regarding the use of Artificial Intelligence (AI) tools. COPE Policy on AI tools can be found at https://publicationethics.org/cope-position-statements/ai-author.

%Authors using AI tools in the writing of a manuscript, production of images or graphical elements of the paper, or in the collection and analysis of data, must be transparent in disclosing in this section how the AI tool was used and which tool was used. Authors are fully responsible for the content of their manuscript, even those parts produced by an AI tool, and are thus liable for any breach of publication ethics. - COPE

%Disclosure instructions

%If there is nothing to disclose, there is no need to add a declaration, otherwise please declare.

%\section*{Use of AI tools declaration}
%The author(s) declare(s) they have used Artificial Intelligence (AI) tools in the creation of this article.
%AI tools used:
%How were the AI tools used?
%Where in the article is the information located?

\section*{Acknowledgements}
The authors wish to thank Miralta Finance Bank S.A. for providing all of the data for the experiments. They also express their gratitude to the reviewers from the journal for their valuable feedback and comments.

\section*{Use of AI tools declaration}
The authors declare they have not used Artificial Intelligence (AI) tools in the creation of this article.

\section*{Conflict of interest}
All authors declare no conflicts of interest in this paper.

%\bibliographystyle{AIMS}
%\bibliography{name}

%For more questions regarding reference style, please refer to the \href{http://www.ncbi.nlm.nih.gov/books/NBK7256/}{Citing Medicine}.

%We encourage authors to submit detailed supplementary, including dataset, document, image, video, software code, protocol, supporting information, table etc, but some large datasets ( $>$ 100 MB) should be deposited in specialized service providers by author. The supplementary should be submitted in a separated file.

\end{document}